\documentclass[twocolumn]{aastex631}

\newcommand{\chandra}{\textit{CXO}}
\newcommand{\fermi}{\textit{Fermi}}

\usepackage{hyperref,amsmath,amssymb}
\usepackage{graphicx}
\numberwithin{equation}{section}
\graphicspath{{Figures/}}

\shorttitle{CXO observations of  13  Fermi LAT Sources}
\shortauthors{Rangelov et al.}

\begin{document}

\title{Chandra X-ray Observatory Observations of 13 Fermi LAT Sources}

\author[0000-0002-9282-5207]{Blagoy Rangelov}
\affil{Department of Physics, Texas State University, 601 University Drive, San Marcos, TX 78666, USA}
\email{rangelov@txstate.edu}

\author{Hui Yang}
\affiliation{Department of Physics, The George Washington University, 725 21st St, NW, Washington, DC 20052, USA}

\author{Brice Williams}
\affiliation{Department of Physics, Texas State University, 601 University Drive, San Marcos, TX 78666, USA}
\affiliation{Department of Physics, Washington State University, Webster Hall 1245, Pullman, WA  99164, USA}

\author{Oleg Kargaltsev}
\affiliation{Department of Physics, The George Washington University, 725 21st St, NW, Washington, DC 20052, USA}

\author{Jeremy Hare}
\affil{Astrophysics Science Division, NASA Goddard Space Flight Center, 8800 Greenbelt Rd, Greenbelt, MD 20771, USA}
\affiliation{Center for Research and Exploration in Space Science and Technology, NASA/GSFC, Greenbelt, Maryland 20771, USA}
\affiliation{The Catholic University of America, 620 Michigan Ave., N.E. Washington, DC 20064, USA}

%\author[0000-0002-7481-5259]{George G. Pavlov}
%\affil{Department of Astronomy \& Astrophysics, Pennsylvania State University, 525 Davey Lab., University Park, PA 16802, USA}

\author{Kean Martinic}
\affiliation{Department of Physics, Texas State University, 601 University Drive, San Marcos, TX 78666, USA}
\affiliation{Department of Physics, Idaho State University, 6921 S. 8th Ave, Pocatello, ID 83209, USA}

\begin{abstract}

In the latest data release from the Fermi Gamma-ray Space Telescope (the 4th Fermi LAT 14\,yr Catalog, or 4FGL), more than 50\% of the Galactic sources are yet to be identified. 
We observed 13 unidentified Fermi LAT sources with the Chandra X-Ray Observatory to explore their nature. We report the results of the classification of X-ray sources in the fields of these $\gamma$-ray sources and discuss the implications for their nature. 
We use multiwavelength (MW) data for a machine-learning  classification, accompanied by a more detailed spectral/variability analysis for brighter sources. 
Eight 4FGL sources have $\gamma$-ray pulsars within their position error ellipses. 
We consider three of these pulsars (PSR J1906+0722, PSR J1105--6037, and PSR J1358--6025) to be detected in X-rays, while PSR J1203--6242 shows a hint of X-ray emission. 
Within the positional uncertainties of three of the 4FGL sources, we detect X-ray sources that may be yet unknown pulsars, depending on the MW association.  In addition to point sources, we discovered two extended sources, one of which is likely to be a bow-shock pulsar-wind nebula associated with PSR J1358--6025. Finally, we classify other X-ray sources detected in these observations and report the most interesting classifications.

\end{abstract}


\section{Introduction} \label{sec:intro}

During the past decade, observations with GeV $\gamma$-ray observatories, such as the Large Area Telescope (LAT) on board the NASA Fermi Gamma-ray Space Telescope (\fermi), have revealed a large number of very high energy sources in the Galactic plane \citep{Acero2015}. Confidently identified sources such as pulsar-wind nebulae (PWNe), supernova remnants (SNRs), active galactic nuclei (AGNs), and X-ray binaries (XRBs) make up only $\approx$38\% of the Galactic ($|b|<5^{\circ}$) GeV $\gamma$-ray sources in the 4th {\it Fermi} LAT 14\,yr Point Source Catalog (4FGL-DR4, hereafter 4FGL; \citealt{2023arXiv230712546B}). 
One of the most commonly identified types of Galactic sources are pulsars. Indeed, there has been a dramatic increase in the number of detected $\gamma$-ray pulsars since the launch of the {\it Fermi}/LAT. In the \emph{EGRET} era, only six $\gamma$-ray pulsars had been discovered, while since  the Fermi LAT launch, the number of known $\gamma$-ray pulsars\footnote{See \url{https://confluence.slac.stanford.edu/display/GLAMCOG/Public+List+of+LAT-Detected+Gamma-Ray+Pulsars} for the latest count.} has increased to 297 (as of 2023 December 2). 
While many pulsars are identified through their $\gamma$-ray and radio pulsations, imaging X-ray observations can reveal young pulsar candidates either through their spectral properties and the absence of multiwavelength (MW) associations, or by resolving extended emission from PWNe. In addition to young pulsars, GeV $\gamma$-ray sources can be associated with pulsars orbiting high-mass stars (high-mass $\gamma$-ray binaries, or HMGBs) or with old recycled pulsars orbiting low-mass stars (redback and black widow systems). Finally, isolated recycled pulsars, which were either kicked out of the binary or were completely ablated by their binary companion, can be sources of GeV $\gamma$-rays. 

In this paper, we report Chandra X-ray Observatory (\chandra) observations of 13 {\it Fermi} LAT sources (listed in Table~\ref{tab:FGL_sample}) that were unidentified  when the CXO observations were proposed (except for PSR J1906+0722). To maximize the probability of finding pulsars, the sources were selected to be close to the Galactic plane ($|b| < 3.5^\circ$) with a LAT detection significance over 15$\sigma$, a LAT position uncertainty $\delta r < 2'$, and a $\gamma$-ray flux (100\,MeV--100\,GeV) $G_{100}>10^{-11}$\,erg\,s$^{-1}$cm$^{-2}$. For the selected sources, we also required the variability index to be $<60$ and the significance of the GeV spectral curvature to be $>5\sigma$. 
The \chandra\ observations and data reduction are described in Section \ref{sec:data}, which is followed by the description of the automated MW classification method in Section \ref{sec:muwclass}.  In Section  \ref{sec:FGL} we discuss the X-ray and MW content of the fields of $\gamma$-ray sources and possible classifications of X-rays sources located within the 95\% confidence level positional uncertainties (PUs) of the 4FGL sources. The two discovered extended X-ray sources are described and discussed in Section \ref{sec:extended}. Other interesting X-ray sources (i.e., capable of producing $\gamma$-rays in principle)  are briefly discussed  in  Section \ref{sec:other-sources}. We present the conclusions in Section \ref{sec:concl}.

%%%%%%%%%%%%%%%%%%%%%%%%%%%%%%%%%%%%%%%%%%%%%%%%%%%%%
%%%%%%%%%%%%%%%%%%%%%%%%%%%%%%%%%%%%%%%%%%%%%%%%%%%%%
\begin{deluxetable*}{ccccccc}
\tabletypesize{\scriptsize}
\tablecaption{Summary of ACIS-I Observations
\label{tab:FGL_sample}}
\tablehead{
\colhead{ObsID} & \colhead{Date} & \colhead{Exposure} & \colhead{FGL Target} & \colhead{$l$} & \colhead{$b$}  & \colhead{$G_{100}$} 
\\
\colhead{} & \colhead{} & \colhead{(ks)} & \colhead{} &\colhead{(deg)} &  \colhead{(deg)} & \colhead{(10$^{-11}$\,cgs)}
}
%\colnumbers
\startdata
20324 & 2018-05-11 & 9.92 & 3FGL J1306.4--6043 & 304.768 & 2.086 & 3.51 \\
20325 & 2018-02-18 & 9.76 & 3FGL J1906.6+0720 & 41.193 & -0.028 & 9.95 \\
20326 & 2018-07-07 & 9.92 & 3FGL J1104.9--6036 & 290.233 & -0.389 & 3.72\\
23586 & 2021-07-10 & 9.93 & 4FGL J1736.1--3422 & 354.31 & -1.17 & 1.9  \\
23587 & 2021-07-24 & 9.92 & 4FGL J1639.3--5146 & 334.25 & -3.33 & 2.7  \\
23588 & 2021-01-28 & 9.92 & 4FGL J1203.9--6242 & 297.52 & -0.34 & 3.8 \\
23589 & 2021-06-20 & 9.93 & 4FGL J0854.8--4504 & 265.61 & -0.03 & 2.3 \\
23590 & 2020-12-14 & 9.95 & 4FGL J2035.0+3632 & 76.6 & -2.34 & 1.2 \\
23591 & 2020-10-25 & 9.91 & 4FGL J2041.1+4736 & 86.1 & 3.45 & 2.5 \\
23592 & 2020-11-29 & 9.92 & 4FGL J1317.5--6316 & 305.86 & -0.56 & 1.7 \\
23593 & 2021-03-28 & 9.92 & 4FGL J1329.9--6108 & 307.56 & 1.38 & 1.4 \\
23594 & 2020-10-10 & 9.92 & 4FGL J0744.0--2525 & 241.35 & -0.74 & 2.0\\
23595 & 2021-02-02 & 10.61 & 4FGL J1358.3--6026 & 311.1 & 1.36 & 3.1 \\
\enddata
\tablecomments{$G_{100}$ is the {\it Fermi} LAT $\gamma$-ray energy flux (100\,MeV--100\,GeV).}%, in units of 10$^{-11}$\,erg\,s$^{-1}$cm$^{-2}$.}%, with uncertainty in parentheses.}
\end{deluxetable*}
%%%%%%%%%%%%%%%%%%%%%%%%%%%%%%%%%%%%%%%%%%%%%%%%%%%%%
%%%%%%%%%%%%%%%%%%%%%%%%%%%%%%%%%%%%%%%%%%%%%%%%%%%%%

%%%%%%%%%%%%%%%%%%%%%%%%%%%%%%%%%%%%%%%%%%%%%%%%%%%%%
\section{Observations and Data Reduction}
\label{sec:data}

We observed all 13 targets over two \chandra\ programs (Table~\ref{tab:FGL_sample}). 
At the time of the earlier program (which included three targets), only the 3rd {\it Fermi} LAT 4\,yr Point Source Catalog \citep[3FGL;][]{2015ApJS..218...23A} was available. All 13 targets were imaged on the ACIS-I array, which was operated in VFAINT mode.

The software Chandra Interactive Analysis of Observations (CIAO), version 4.15, and the Chandra Calibration Data Base (CALDB), version 4.10.4, were used to reduce and analyze the data \footnote{\href{http://cxc.harvard.edu/ciao/}{http://cxc.harvard.edu/ciao/}}. We used the CIAO Mexican-hat wavelet source detection routine \texttt{wavdetect} \citep{Freeman2002} to create source lists. Wavelets of 2, 4, 8, and 16 pixels and a detection threshold of $10^{-6}$ were used. 

We used an exposure-time-weighted average point-spread function (PSF) map and adhered to standard CIAO procedures\footnote{\url{http://cxc.harvard.edu/ciao/threads/}} to determine the source fluxes, spectral properties, and variability. 
Each observation was run through the CIAO tool \texttt{srcflux} with the coordinates obtained by \texttt{wavdetect}. The data were filtered in three energy bands, 0.5--1.2\,keV (soft), 1.2--2.0\,keV (medium), and 2.0--7.0\,keV (hard), and were limited to the energy range between 0.5 and 7.0 keV.\footnote{Due to the increasing buildup of a contaminating layer on the ACIS detector and uncertainties in the ACIS quantum efficiency contamination model near the O-K edge at 0.535\,keV and F-K edge at 0.688\,keV (see \url{https://cxc.cfa.harvard.edu/ciao/why/acisqecontamN0010.html}), we used the 0.7--1.2\,keV energy range for detection and then converted the energy fluxes into 0.5--1.2\,keV using scaling factors calculated by assuming a power-law (PL) spectrum with a photon index $\Gamma=2.0$.} Whenever specified, the uncertainties of the best-fit spectral parameters and ACIS-I fluxes are given at 68\% level.

The absolute astrometry of the X-ray images was corrected for by aligning the X-ray sources to reference sources from the Gaia DR3 catalog \citep{2023A&A...674A...1G}, using the method similar to that used by the Chandra Source Catalog team\footnote{\url{https://cxc.cfa.harvard.edu/csc/memos/files/Martinez-Galarza_stack_astrometry_translation_spec_doc.pdf}}.
The details of the astrometric correction are described in detail in \cite{2024ApJ...971..180Y}. %H. Yang et al. (2023, in preparation). 

%%%%%%%%%%%%%%%%%%%%%%%%%%%%%%%%%%%%%%%%%%%%%%%%%%%%%
\section{Automated Multiwavelength Classification}
\label{sec:muwclass}

The multiwavelength machine-learning classification (MUWCLASS) pipeline developed by \cite{2022ApJ...941..104Y} was used to perform an automated classification of X-ray sources detected in 
all 13 4FGL fields that are analyzed in this paper. The pipeline trains a random forest machine-learning (ML) model with a training data set (TD) of $\sim3000$ sources of literature-verified classes from the Chandra Source Catalog version 2.0 (CSCv2.0), presented in \cite{2022ApJ...941..104Y} (with minor updates described in \citealt{2024ApJ...971..180Y}), including AGNs, low-mass stars (LM-STARs), high-mass stars (HM-STARs), young stellar objects (YSOs), cataclysmic variables (CVs), high-mass XRBs (HMXBs), low-mass XRBs (LMXBs)\footnote{This class also includes nonaccreting XRBs, such as wide-orbit binaries with millisecond pulsars (MSPs), as well as redback and black widow systems.}, and neutron stars (NSs). In addition to the X-ray properties (fluxes, hardness ratios, and X-ray variability measures), each source is characterized by its 
optical/near-infrared (NIR)/infrared (IR) properties from the Gaia DR3 \citep{2023A&A...674A...1G}, 2MASS \citep{2006AJ....131.1163S}, AllWISE \citep{2014yCat.2328....0C}, and CatWISE2020 \citep{2021ApJS..253....8M} catalogs using the probabilistic cross-matching algorithm NWAY (\citealt{2018MNRAS.473.4937S}). 
The NWAY algorithm provides $p\_{\rm any}$\footnote{We kept the original notation from \cite{2018MNRAS.473.4937S} here.}, the probability of having a true counterpart among all considered associations, and $p\_{\rm i}$, the probability of each ($i$th) counterpart (including the case of having no association at all) being the right counterpart. 
Based on $p\_{\rm any}$ and $p\_{\rm i}$, we calculate the chance coincidence probability $P_{\rm c}=1-p\_{\rm any}$ and the association probability $P_{\rm i}=p\_{\rm i}\times p\_{\rm any}$ for each possible association. $P_{\rm i}$ and $P_{\rm c}$ are calculated per specific catalog, and they are indicated by a letter (G for Gaia, C for CatWISE2020, T for 2MASS, A for AllWISE), followed by the values of $P_{\rm i}$ and $P_{\rm c}$ in Tables\,\ref{tab:X-ray-Class-errorellipse} and \ref{tab:X-ray-Class}. For X-ray sources for which no counterpart was the only possibility,  $P_{\rm i}$ and $P_{\rm c}$ were set to 1 and 0, which are different from the probabilities of lacking any counterparts for the X-ray sources having some possible associations, for which we left $P_{\rm i}$ and $P_{\rm c}$ as empty. 

We also implemented an oversampling method to produce synthetic sources by sampling (optical) extinction and (X-ray) absorption values from the distributions of those values for the TD sources and applying them to the less densely populated (excluding AGN) classes. This oversampling is more realistic (physical) than other algorithms (e.g., SMOTE; \citealt{2011arXiv1106.1813C}, which was used in \citealt{2022ApJ...941..104Y}), and it better reproduces (due to the additional extinction and absorption) a fainter population of sources  that are more consistent (in terms of fluxes) with the unidentified sources that we classify. Detailed descriptions of the updated TD, the cross-matching method, the oversampling method, and the evaluation of the performance of the updated MUWCLASS pipeline are presented in \cite{2024ApJ...971..180Y}. %H. Yang et al. (2023, in preparation). 

\begin{figure*}
\begin{center}
\includegraphics[width=0.8\columnwidth]{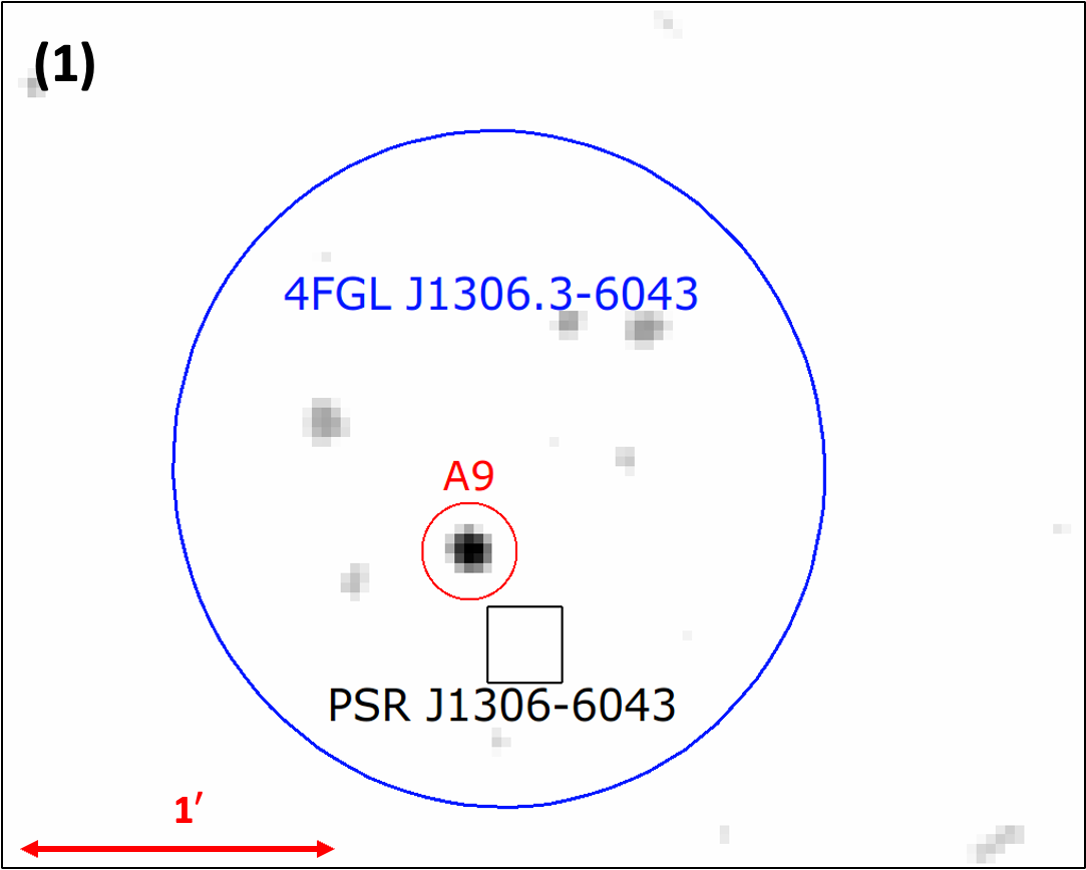}
\includegraphics[width=0.8\columnwidth]{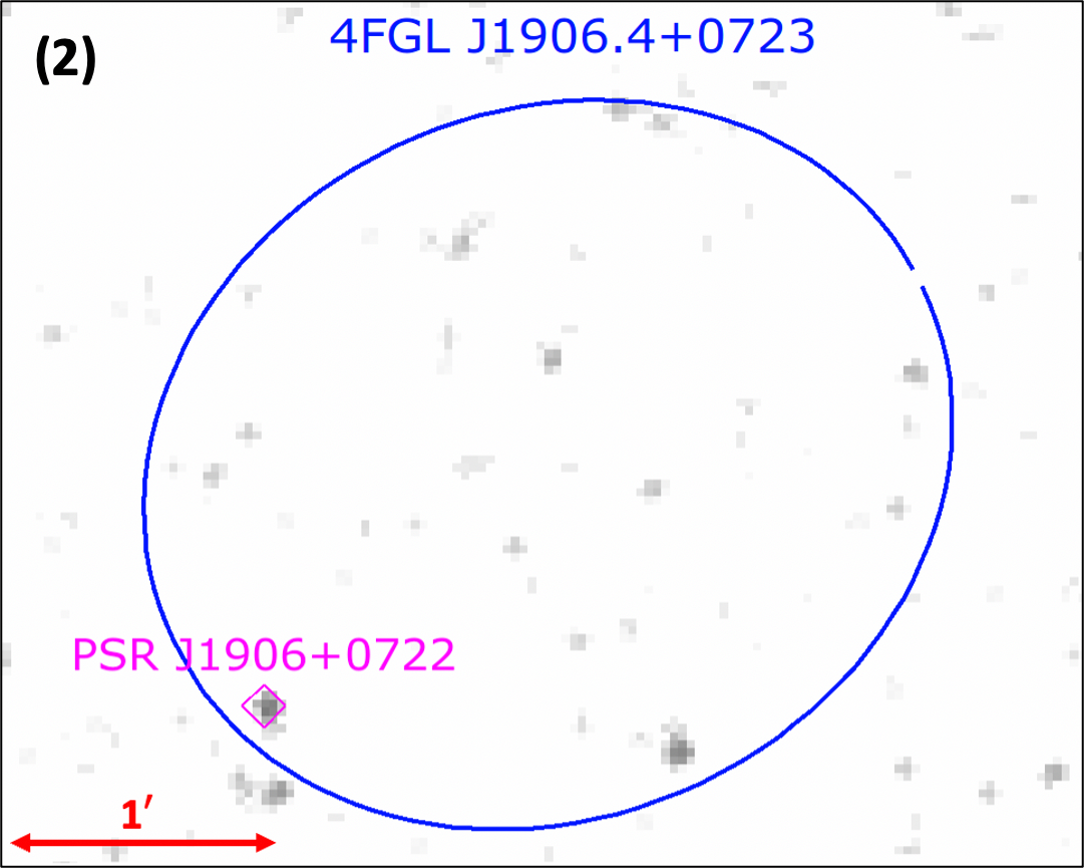}
\includegraphics[width=0.8\columnwidth]{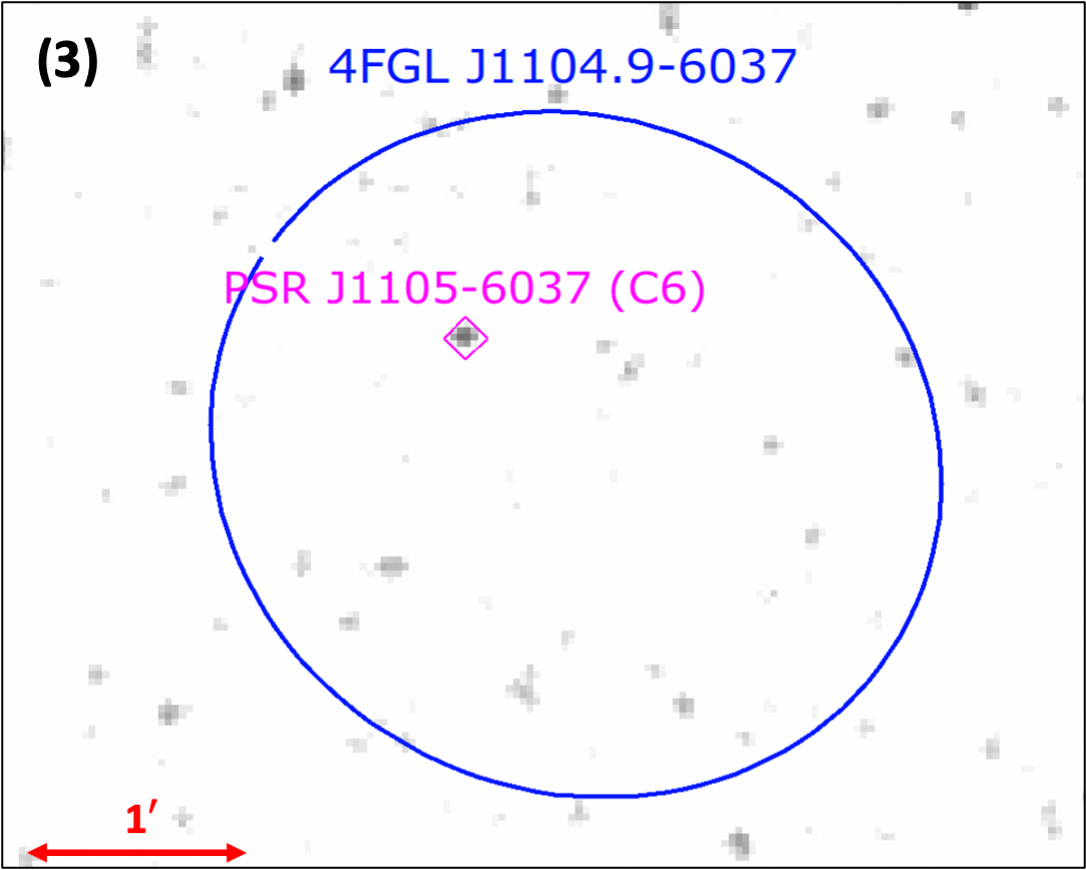}
\includegraphics[width=0.8\columnwidth]{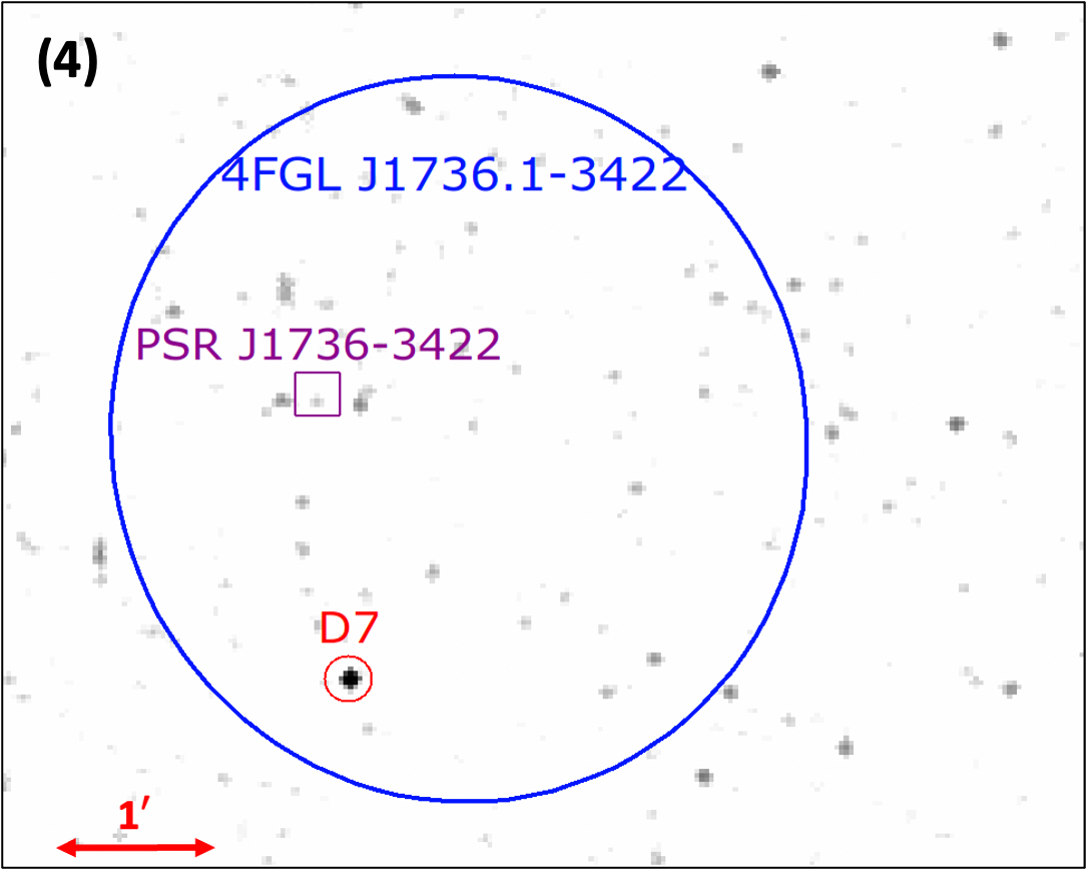}
\includegraphics[width=0.8\columnwidth]{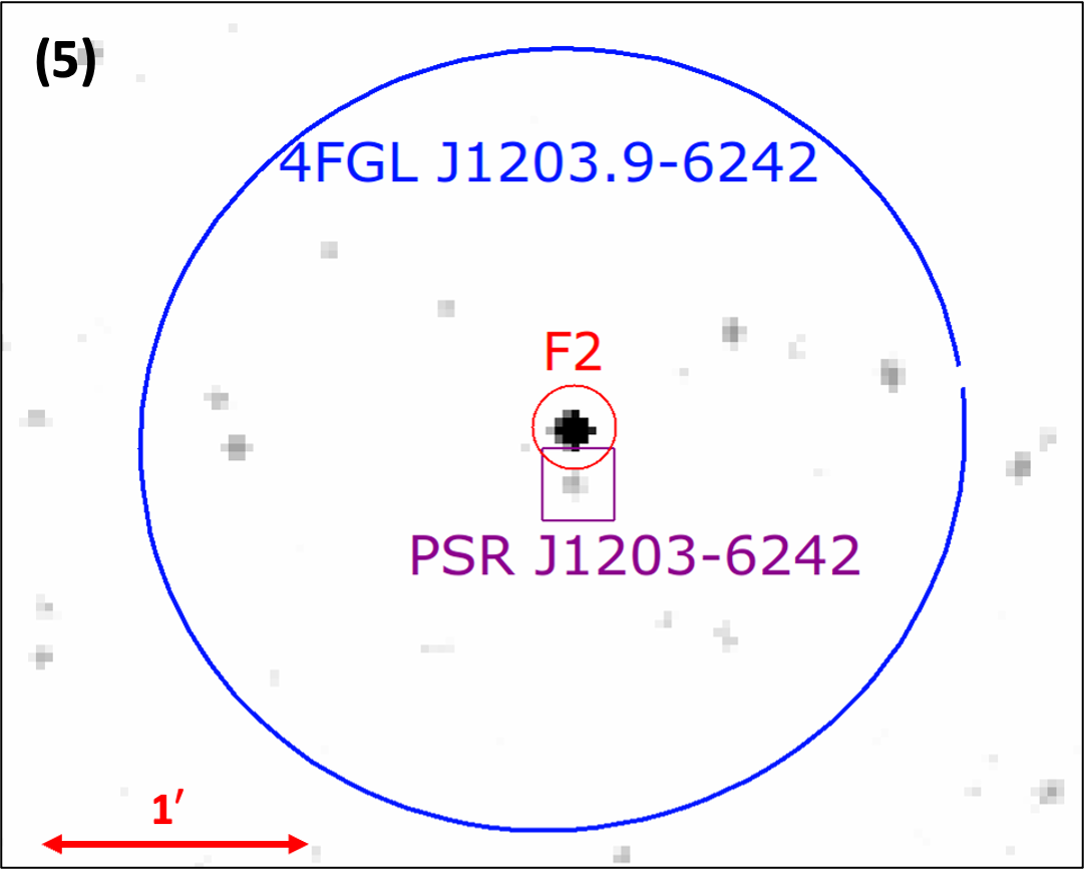}
\includegraphics[width=0.8\columnwidth]{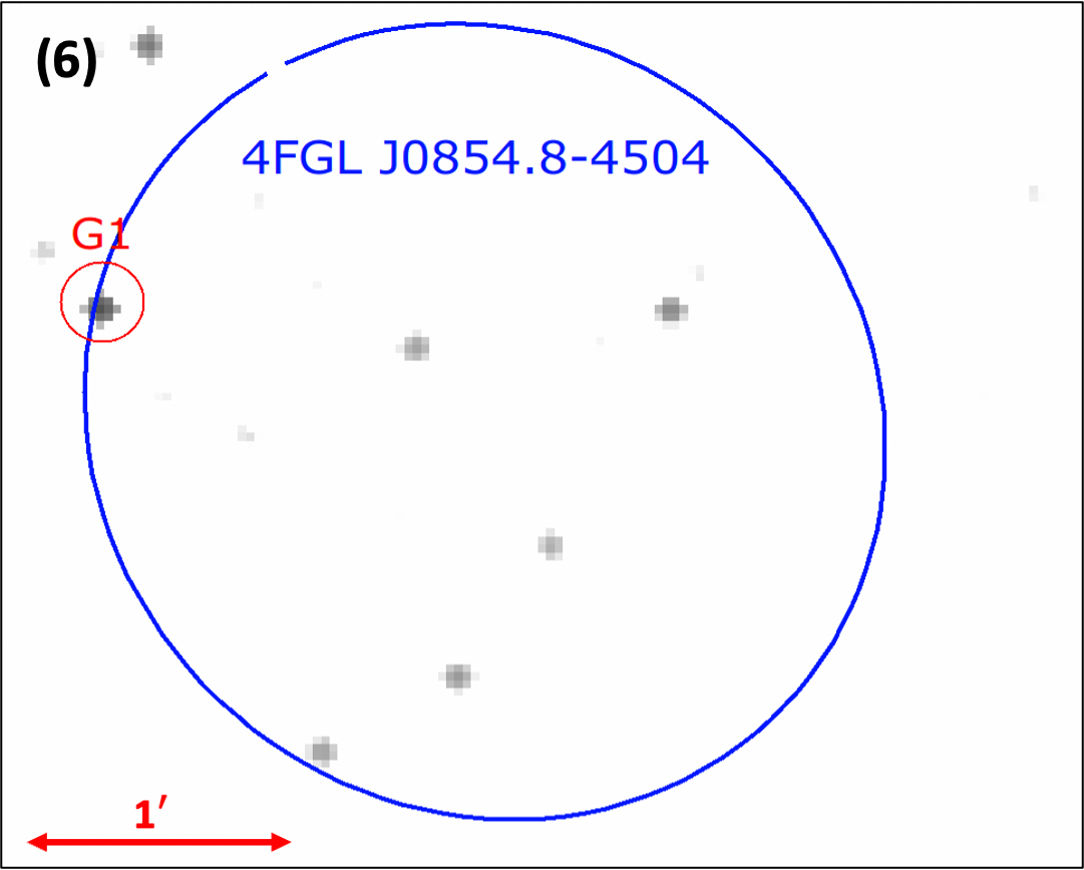}
\includegraphics[width=0.8\columnwidth]{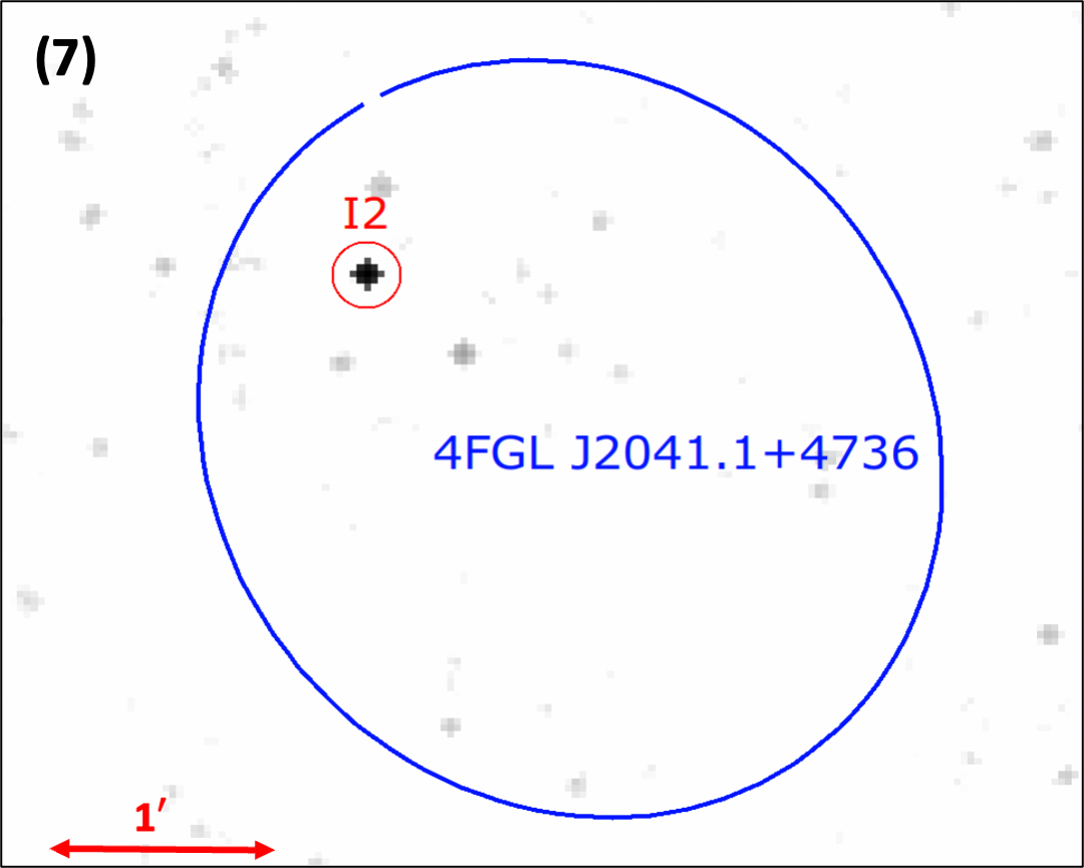}
\includegraphics[width=0.8\columnwidth]{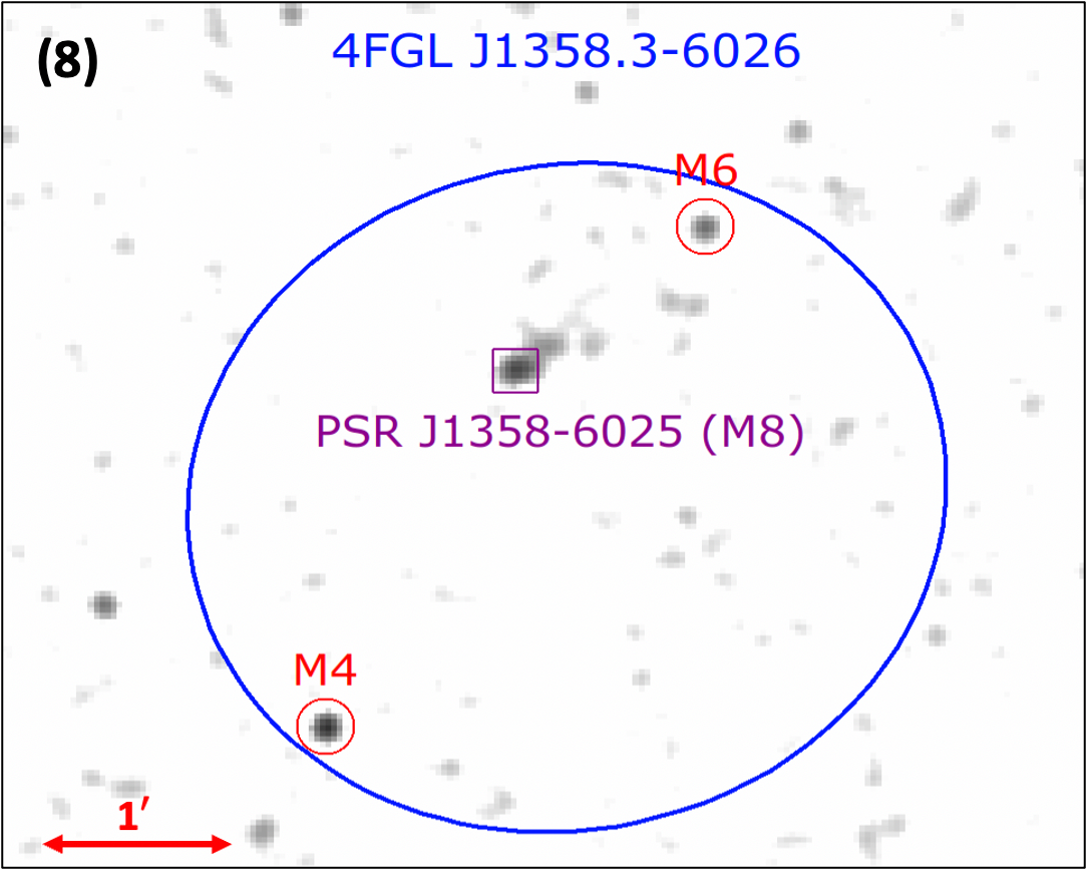}
\caption{
Error ellipses (in blue) of eight 4FGL sources with X-ray sources and/or $\gamma$-ray pulsars detected inside. The red circles indicate the detected X-ray sources. The magenta diamonds indicate the accurately (with subarcsecond precision) known positions of $\gamma$-ray pulsars from $\gamma$-ray timing solutions. The purple and black squares indicate the locations of the $\gamma$-ray pulsars from the 4FGL-DR4 catalog and MMGPS, for which the uncertainties are not yet published. 
} 
\label{fig:FGL-fields-zoomin}
\end{center}

\end{figure*}

\begin{figure*}
\begin{center}
\includegraphics[width=0.8\columnwidth]{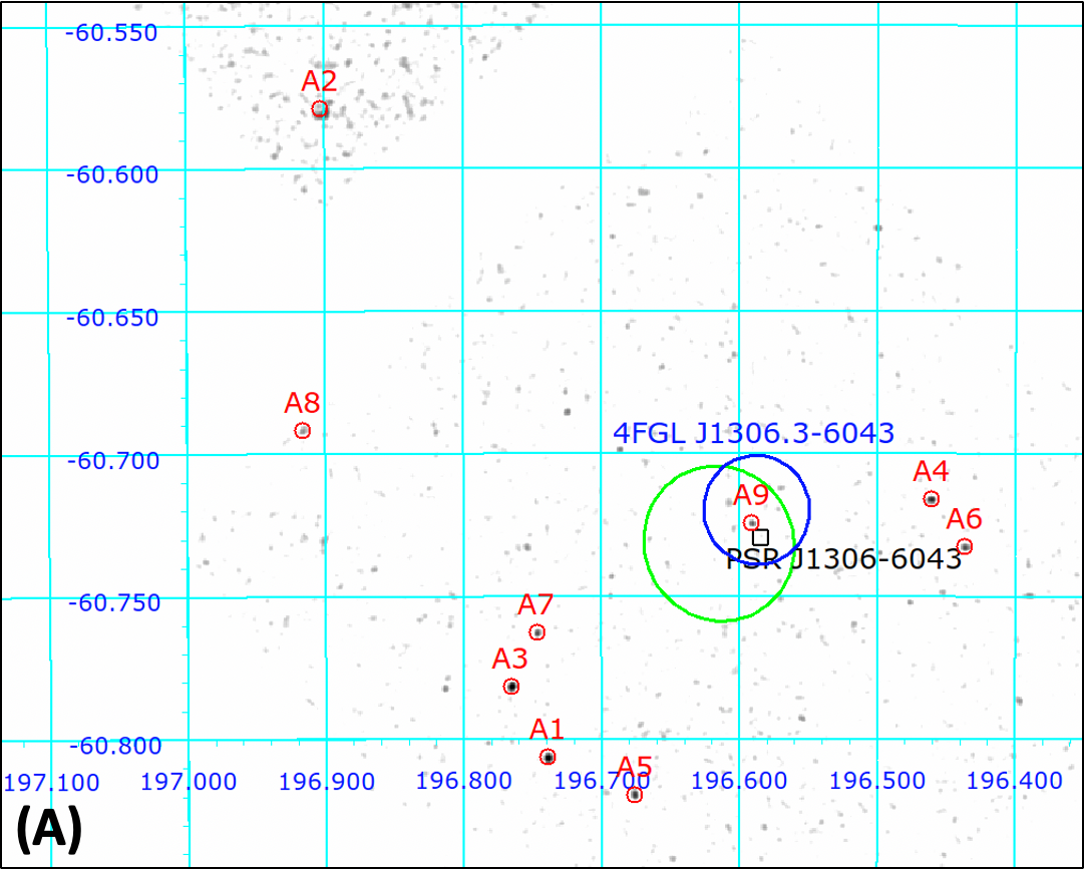}
\includegraphics[width=0.8\columnwidth]{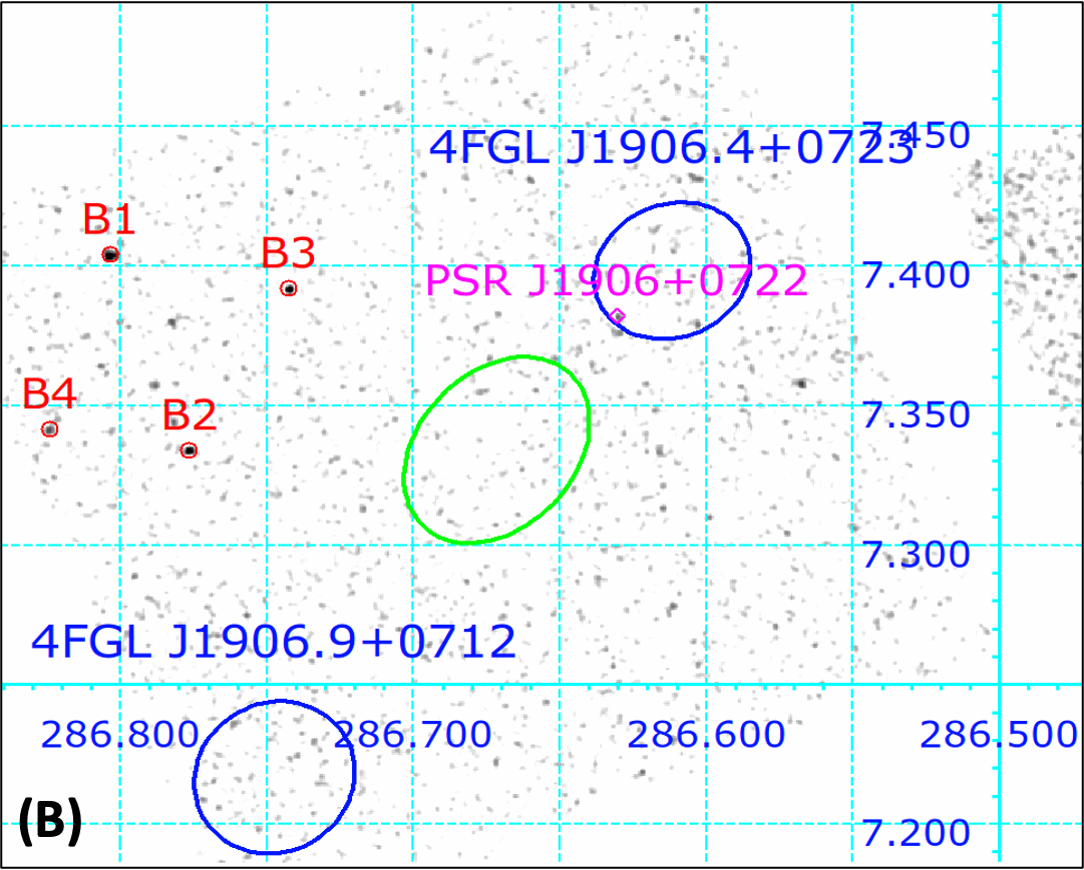}
\includegraphics[width=0.8\columnwidth]{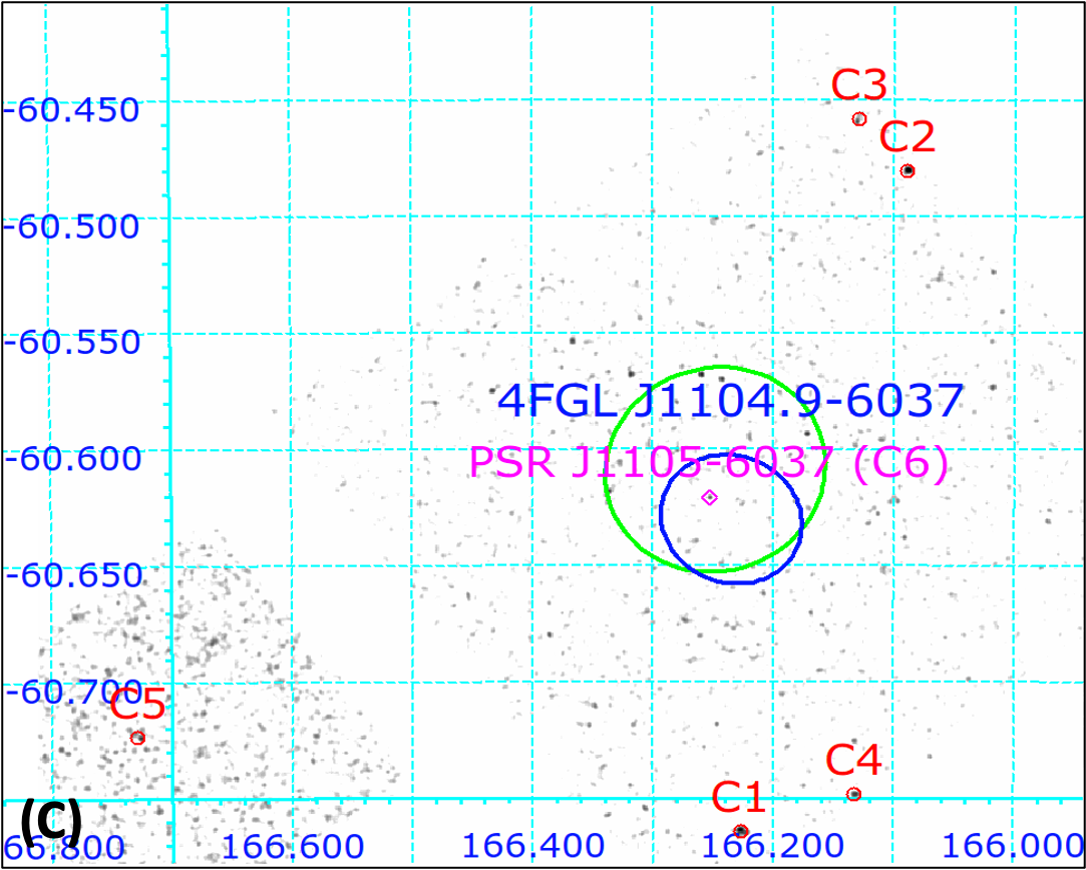}
\includegraphics[width=0.8\columnwidth]{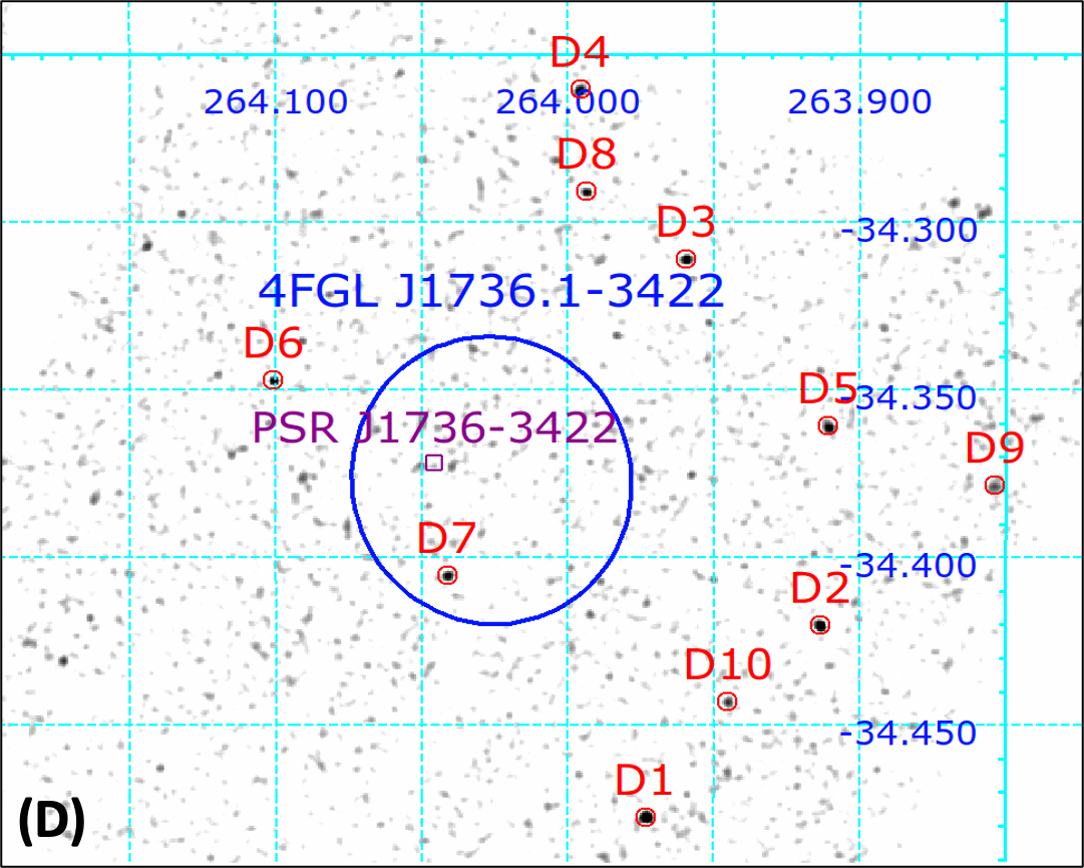}
\includegraphics[width=0.8\columnwidth]{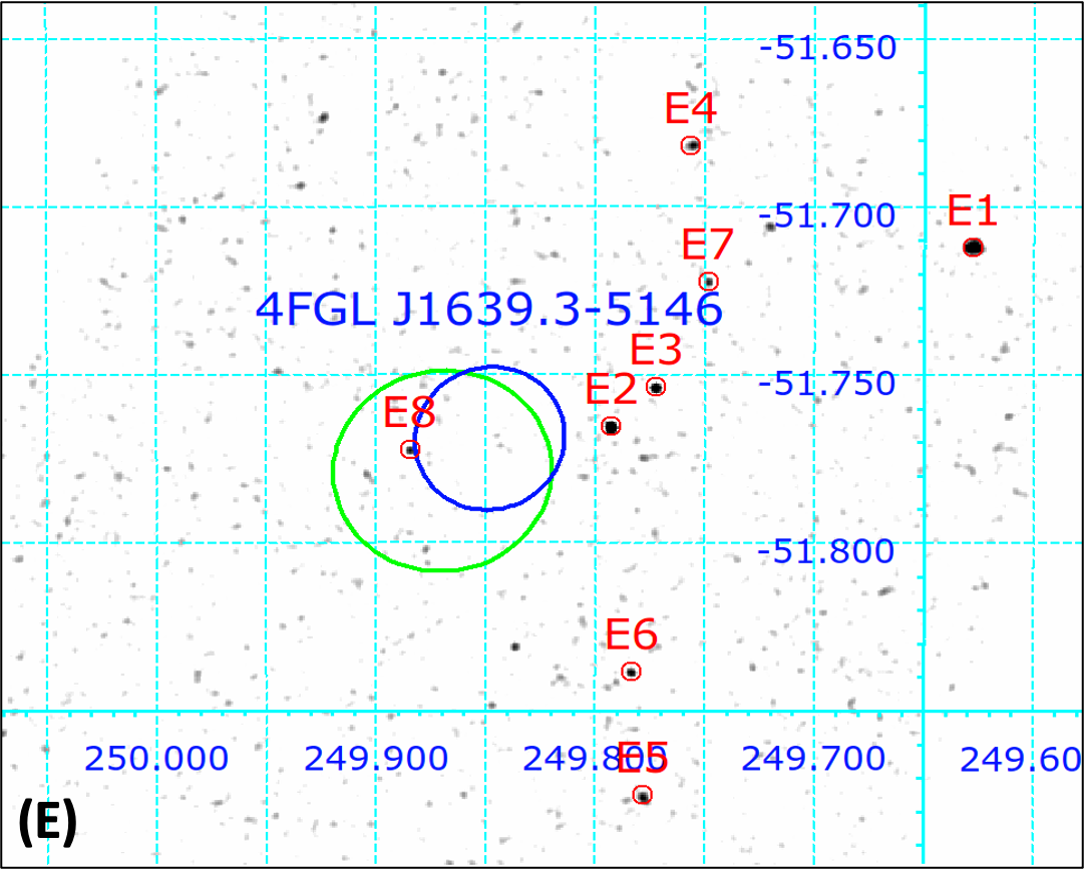}
\includegraphics[width=0.8\columnwidth]{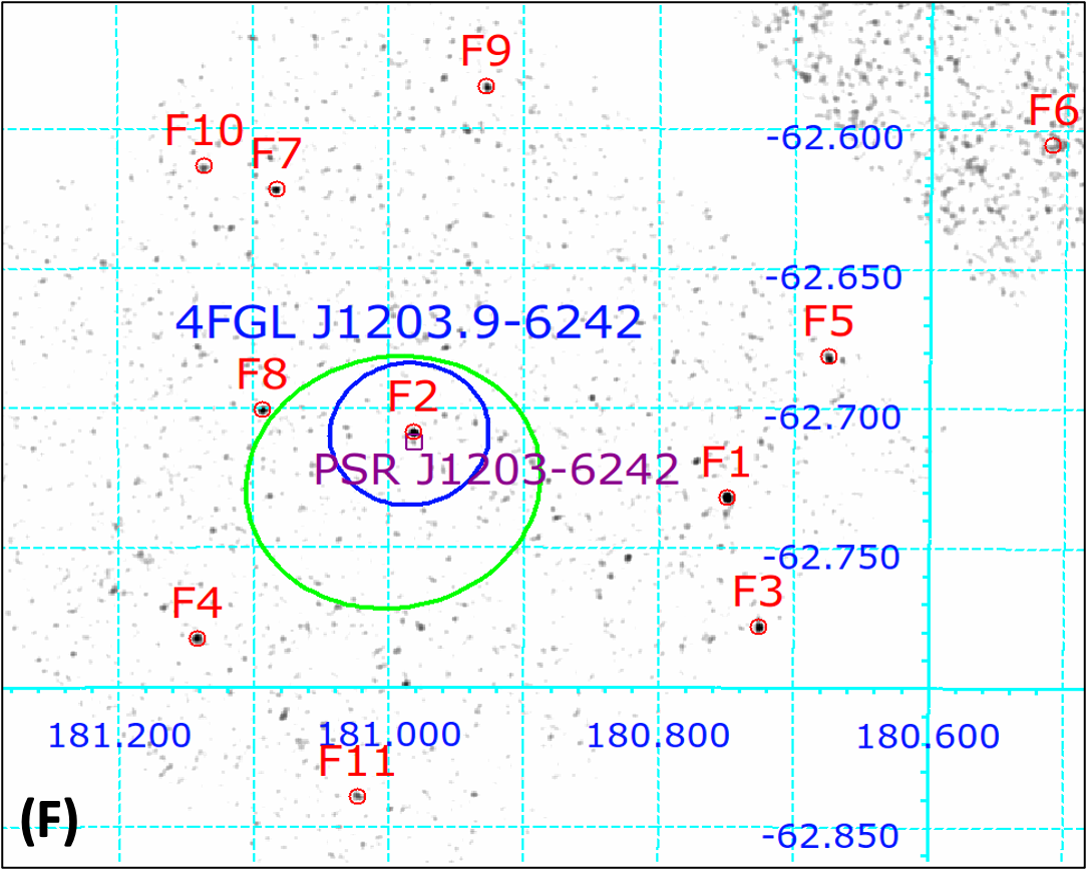}
\caption{
First 6 of 13 Fermi fields with 4FGL-DR4 error ellipses shown in blue, and the 3FGL error ellipses are shown in green (if available). 
The red circles, magenta diamonds, and purple and black squares are the same as in Figure \ref{fig:FGL-fields-zoomin}.
} 
\label{fig:FGL-fields-1}
\end{center}
\end{figure*}

\begin{figure*}
\begin{center}
\includegraphics[width=0.8\columnwidth]{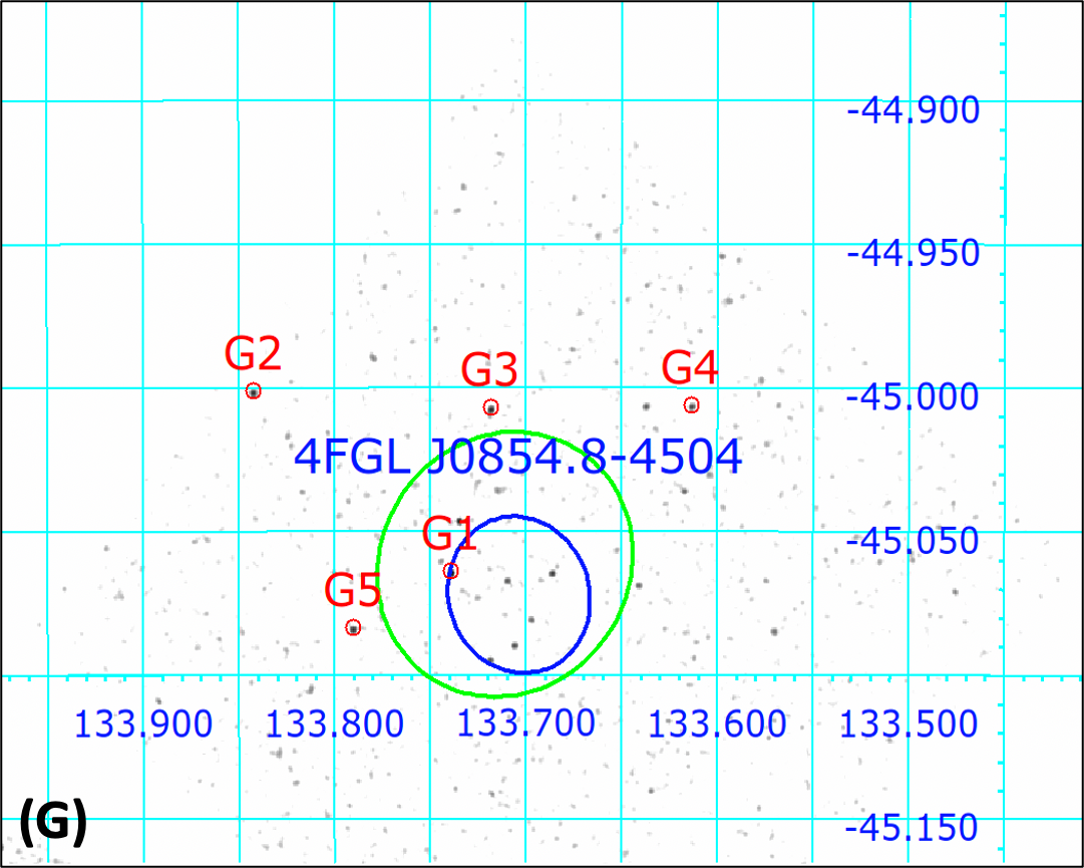}
\includegraphics[width=0.8\columnwidth]{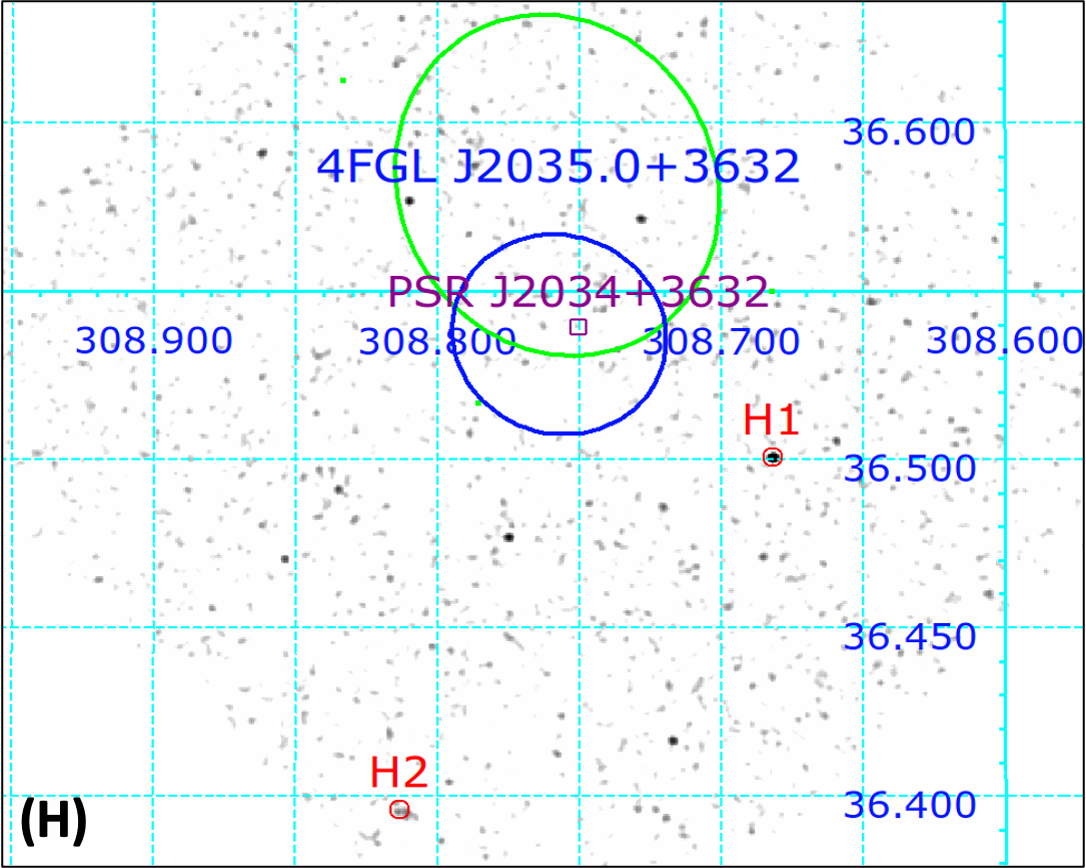}
\includegraphics[width=0.8\columnwidth]{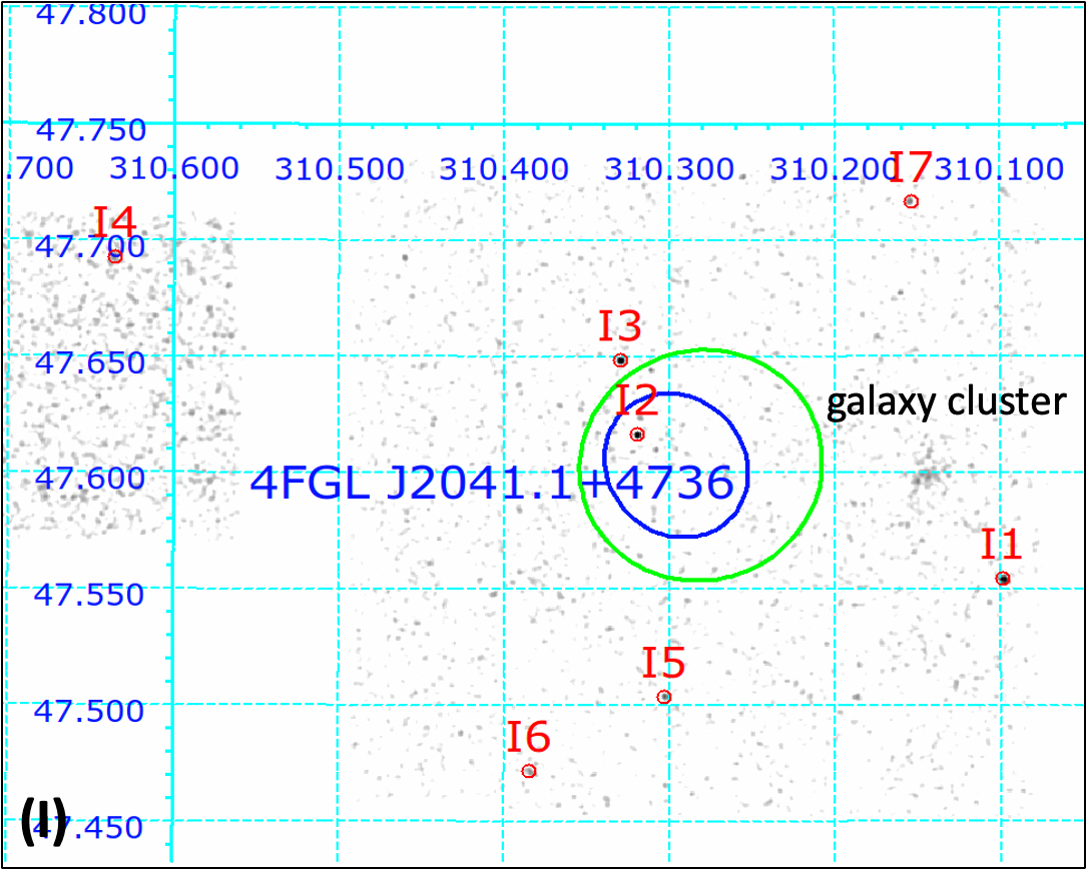}
\includegraphics[width=0.8\columnwidth]{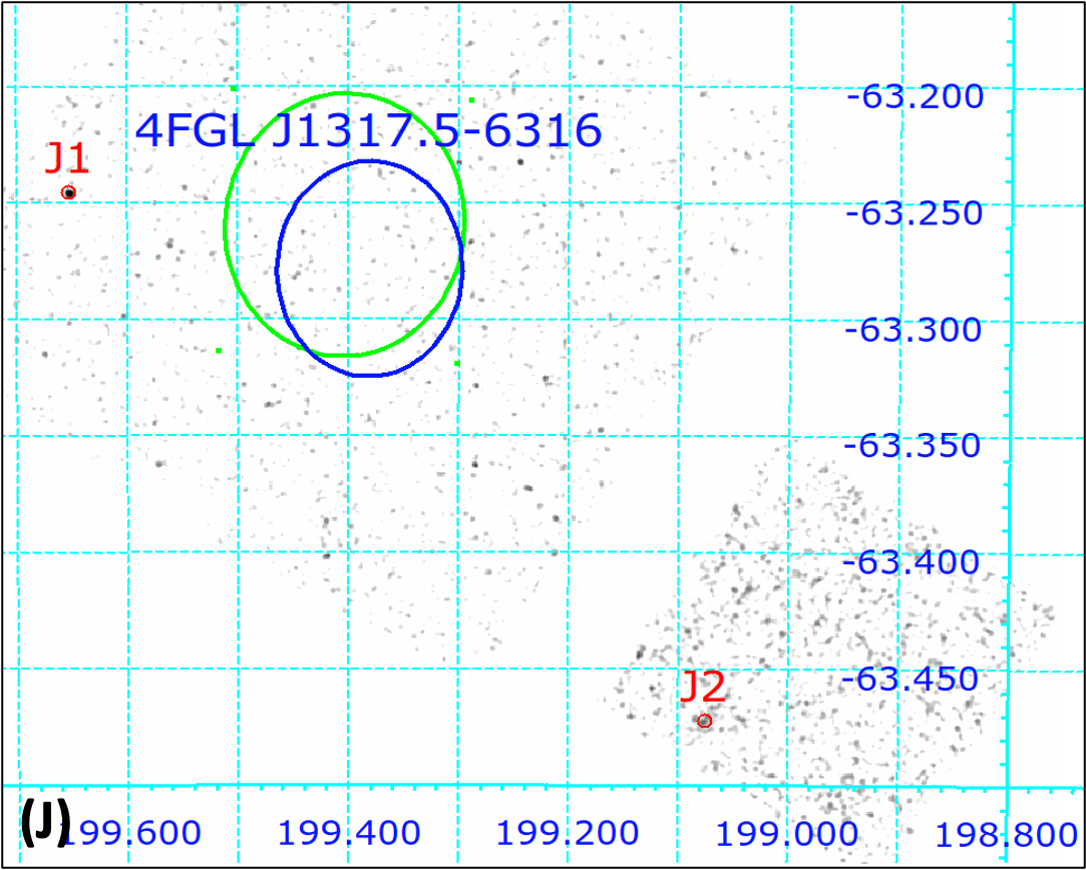}
\includegraphics[width=0.8\columnwidth]{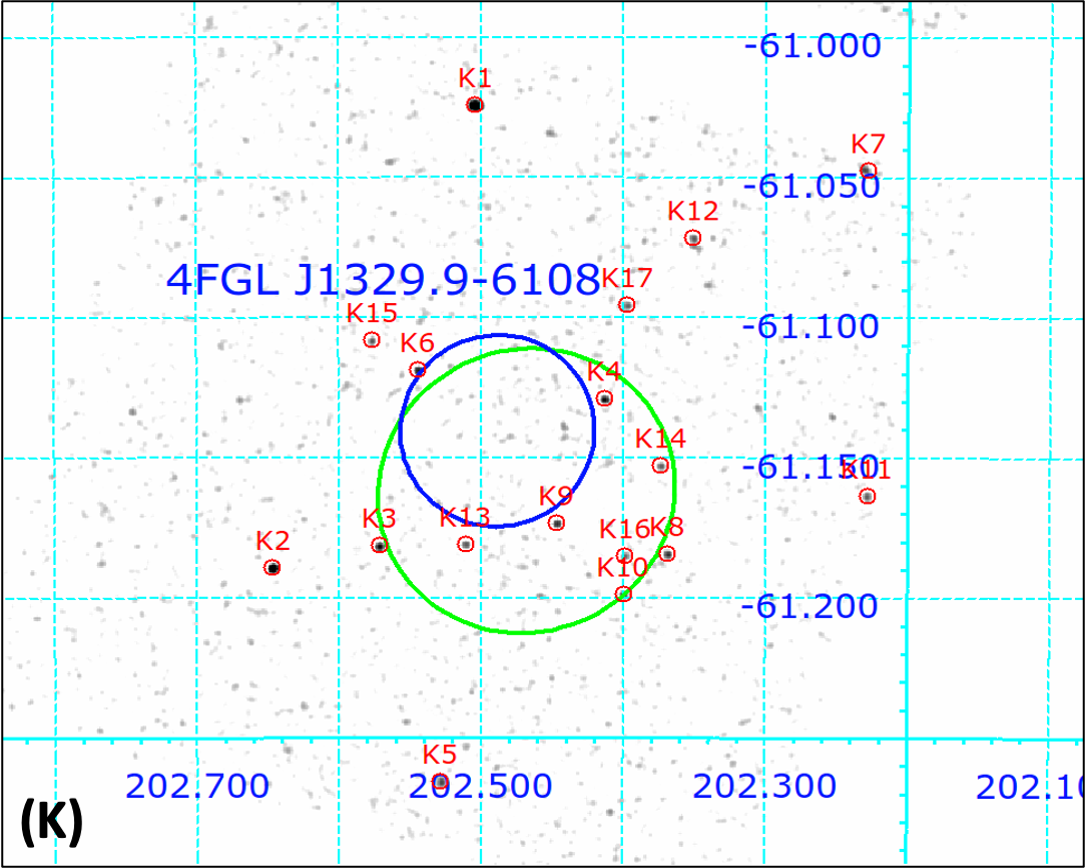}
\includegraphics[width=0.8\columnwidth]{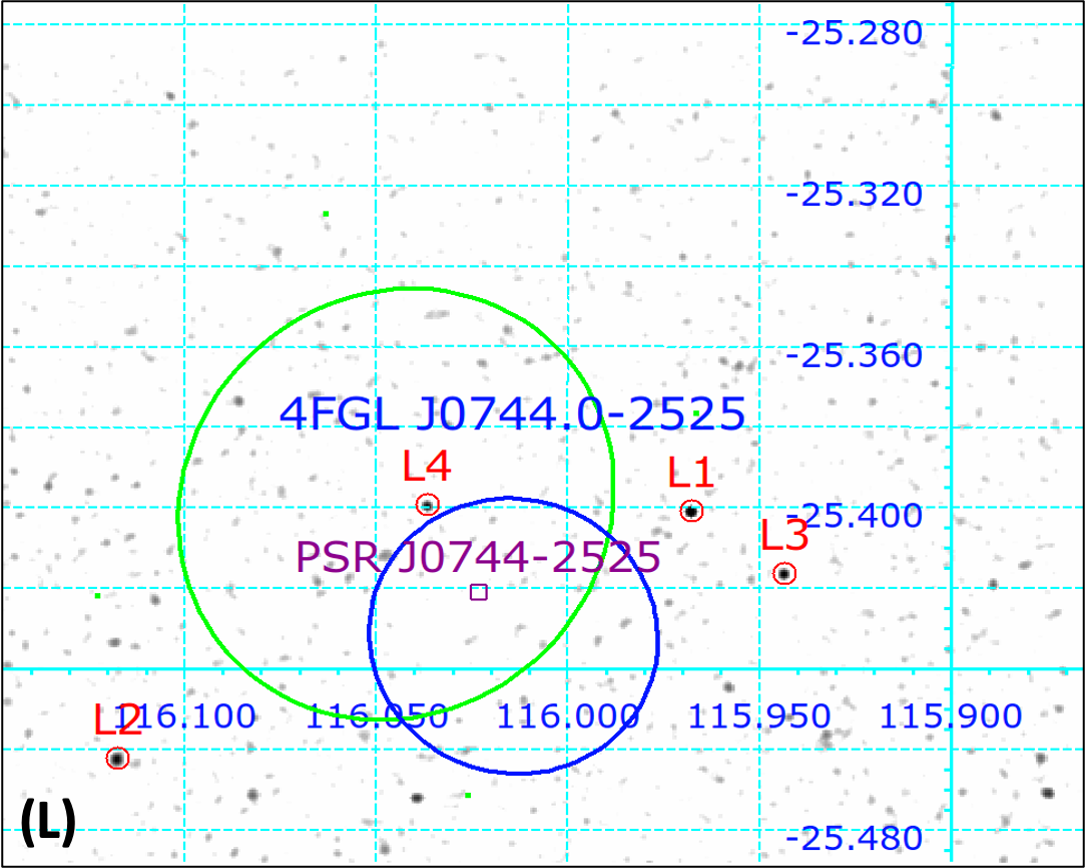}
\includegraphics[width=0.8\columnwidth]{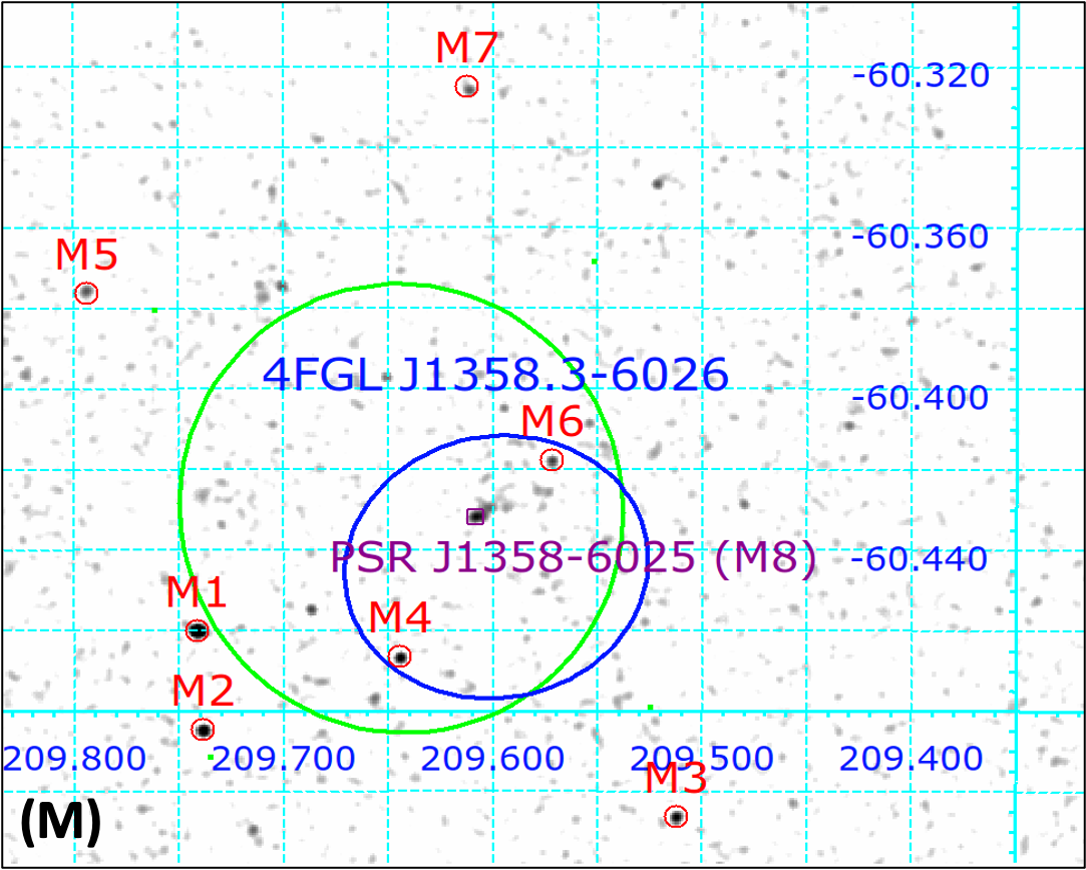}
\caption{
The following 7 of 13 Fermi fields. The markers are the same as in Figure\,\ref{fig:FGL-fields-1}. One extended source  is marked as galaxy cluster in panel (I), and the other extended source is marked as PSR J1358--6025 (M8) in panel (M).
} 
\label{fig:FGL-fields-2}
\end{center}
\end{figure*}

Below, we mostly consider classifications for sources with an X-ray detection significance  $\ge 3$ (according to {\tt wavdetect}), leaving 91 sources in all 13 4FGL fields. We added only two sources that are below this threshold because they are coincident with the locations of discovered $\gamma$-ray pulsars  (see Figure~\ref{fig:FGL-fields-zoomin}; \citealt{2015ApJ...809L...2C,2017ApJ...834..106C}) and that are hence more likely to be real sources than an arbitrary subthreshold detection. For the classifications, we  consider multiple possible associations when more than one counterpart matched the X-ray source within its PU (at the 95\% confidence level), including the possibility of no counterpart. 
For the same X-ray source (uniquely identified by its CXO name), the cases of multiple associations are distinguished by the association names (A-name) given in the second column of Tables\,\ref{tab:X-ray-Class-errorellipse} and \ref{tab:X-ray-Class}, where a dash indicates a specific association (the number after the dash indexes  different associations), while the name without a dash corresponds to a case without a counterpart. 
When all possible associations are accounted for, 29 sources are left without any counterpart, 35 sources have one counterpart, 14 sources have 2 counterparts, and the remaining 15 sources have $>2$ counterparts.
The X-ray sources with many possible counterparts typically have large PUs (mostly because they were imaged at large off-axis angles) and are difficult to classify confidently.
For this reason, we only discuss the classifications for sources with PUs$<=3\arcsec$ below and report them in Tables \ref{tab:X-ray-Class-errorellipse} and \ref{tab:X-ray-Class}. However, the X-ray  sources with larger PUs are still shown in Figure~\ref{fig:FGL-fields-1} and \ref{fig:FGL-fields-2}.

%%%%%%%%%%%%%%%%%%%%%%%%%%%%%%%%%%%%%%%%%%%%%%%%%%%%%
\section{Results and Discussion}
\label{sec:results}

Below, we first discuss each 4FGL source field individually, considering X-ray sources located within the PUs of the 4FGL sources and attempting to identify the nature of the GeV source. After this, we  separately consider two extended X-ray sources and the most interesting sources lying outside of the PUs of the 4FGL sources. 

\subsection{Counterparts of Fermi LAT Sources}
\label{sec:FGL}

In the 13 4FGL fields, eight LAT sources have been associated with $\gamma$-ray pulsars (discovered after our \chandra\ programs were approved) that were detected in the third Fermi LAT catalog of $\gamma$-ray pulsars (3PC; see Table\,\ref{tab:4FGLDR3-NSs}; \citealt{2015ApJ...809L...2C,2017ApJ...834..106C,2023arXiv230711132S}). Two of the $\gamma$-ray pulsars, PSR J1906+0722 and PSR J1105--6037, have been relatively well studied, and their positions are known very precisely \citep{2015ApJ...809L...2C,2017ApJ...834..106C}. For the other six, only arcminute-scale localizations are available from the 3PC. However, the localizations of these six pulsars are listed in the 4FGL-DR4 catalog with a much  higher accuracy (probably based on unpublished information from the 3PC group), and we use these locations for the rest of the paper, except for PSR J1306--6043, for which we use the original position from the radio observations \citep{2023MNRAS.524.1291P} . 

We also checked whether X-ray sources were detected within the $\gamma$-ray PUs in the Swift X-ray Telescope (XRT) survey of unassociated Fermi LAT sources\footnote{\url{https://www.swift.psu.edu/unassociated/}} \citep{2013ApJS..207...28S} and in the  Living Swift XRT Point Source Catalogue \cite[LSXPS;][]{2023MNRAS.518..174E}. 

\subsubsection{4FGL J1306.3--6043}

3FGL J1306.4--6043 (which later became 4FGL J1306.3--6043) has been
associated with the radio pulsar PSR J1306$-$6043 discovered in the MPIfR-MeerKAT Galactic Plane Survey \citep[MMGPS;][]{2023MNRAS.524.1291P}, which has also been detected in $\gamma$-rays from the 3PC \citep{2023arXiv230711132S}. It is an MSP with a spin period of 5.67\,ms, which lives in a potential MSP helium white dwarf system with an orbital period of $\approx$\,86\,d \citep{2023MNRAS.524.1291P}. No X-ray source is seen in the ACIS-I
image within 10$''$ from the position of the radio pulsar.

Within the 95\% PU of the LAT source, only a single source, CXO J130621.8--604328 (A9), is detected in the ACIS image at $\approx3\sigma$ significance (see panel 1 in  Figure \ref{fig:FGL-fields-zoomin}). Because A9 is only $\sim23\arcsec$ away from PSR J1306$–$6043, it might be associated with the MSP if the (unpublished) uncertainty of the pulsar position is large enough. This X-ray source likely has an optical/NIR/IR  counterpart  (Gaia DR3 6055821050370180608, 2MASS 13062177--6043281 and CatWISE2020 J130621.76--604328.1), which we refer to as A9-1 when considering it as an association. 
The MUWCLASS pipeline has classified A9 as an NS with a classification confidence threshold (hereafter CT\footnote{The higher the CT, the more confident the classification. We usually set CT=2 as a cut for the confident classifications \citep{2022ApJ...941..104Y}.}) =1.1 
without the optical-IR counterpart, while A9-1 is classified as a YSO with CT=2.2.

Given that the positional offset of the Gaia source (0.12$''$) is much smaller than the PU of the X-ray source (1.1$''$), we estimate the association probability to be high ($P_{\rm i}=83\%$, based on matching to the Gaia DR3 catalog), and hence, the latter classification is more likely.

\subsubsection{3FGL J1906.6+0720}

In the 4FGL-DR4 catalog, the observed 3FGL J1906.6+0720 source is resolved into two distinct $\gamma$-ray sources: 4FGL J1906.4+0723 and 4FGL J1906.9+0712. The former, 4FGL J1906.4+0723, is associated with the $\gamma$-ray pulsar PSR J1906+0722, whose pulsations were found by the 3PC \citep{2015ApJ...809L...2C,2023arXiv230711132S}. Observations with the Swift XRT provided an upper limit for the unabsorbed X-ray (0.5--10\,keV) flux of $2\times10^{-13}$\,erg\,cm$^{-2}$s$^{-1}$ \citep{2015ApJ...809L...2C}.

There is no compelling evidence that PSR J1906+0722 is associated with any of the four known SNRs located within 1\,$^{\circ}$ of the pulsar timing position. The closest SNR, G41.1--0.3, is likely a Type Ia SNR \citep{2015ApJ...809L...2C}. This makes it an unlikely birthplace for a pulsar. Due to their proximity, the association between the G41.1--0.3 and the second 4FGL source, 4FGL J1906.9+0712 has a much higher probability than the associations with G41.1--0.3 and 4FGL J1906.4+0723.

Although \texttt{wavdetect} did not report any significant (S/N$>3$) X-ray sources within the 95\% PU ellipse of 4FGL J1906.4+0723, a faint and possibly slightly extended emission is discernible at the pulsar location determined from the $\gamma$-ray timing (see panel 2 in Figure \ref{fig:FGL-fields-zoomin}). 
Because the observed X-ray excess is precisely coincident with the accurately known $\gamma$-ray timing position, we consider it unlikely to be a noise fluctuation. The somewhat extended nature of the emission hints at the possibility that this is a PWN. 
Running \texttt{srcflux} at this position yields a flux of $(3.4_{-1.4}^{+1.9})\times10^{-14}$\,erg\,cm$^{-2}$\,s$^{-1}$ within the 90\% encircled PSF aperture ($r=2.7\arcsec$).

We did not detect any significant ($\ge3\sigma$) X-ray sources in the ACIS-I image within the 4FGL J1906.9+0712 PU.

\subsubsection{4FGL J1104.9--6037}

4FGL J1104.9--6037 has been associated with the $\gamma$-ray pulsar PSR J1105--6037, identified 
by the 3PC \citep{2017ApJ...834..106C,2023arXiv230711132S}. A potential X-ray counterpart, 1SXPS J110500.3--603713, was reported in \cite{2018ApJ...854...99W}, with an observed flux of $2.5^{+1.5}_{-1.1}\times10^{-13}$\,erg\,cm$^{-2}$\,s$^{-1}$ in the 0.3--10\,keV, while the LSXPS has only provided a 3$\sigma$ upper limit on the observed flux of $\sim2\times10^{-13}$\,erg\,cm$^{-2}$\,s$^{-1}$. 

In the ACIS-I image, a faint source ($\sim1.7\sigma$), CXO J110500.5--603715 (C6), coincides with the position of PSR J1105--6037  (see panel 3 in Figure~\ref{fig:FGL-fields-zoomin}). The source does not pass the significance threshold and lies near the ACIS-I chip gap\footnote{There is a gap of 11$\arcsec$ (22 pixels) between every two neighboring ACIS-I charge-coupled device (CCD) chips. For a more technical description of the ACIS instrument, see \url{https://cxc.harvard.edu/cal/Acis/index.html}.}. 
Because the observed X-ray excess is precisely coincident with the accurately known $\gamma$-ray timing position, we consider it unlikely that this is a noise fluctuation. 
The observed broadband flux of $\sim2.5\times10^{-15}$\,erg\,cm$^{-2}$\,s$^{-1}$ (in 0.5--7\,keV) 
primarily comes from the medium band (1.2--2.0\,keV).
The flux difference between the CXO and Swift observations suggests that the X-ray source might be variable, which is unexpected for an isolated pulsar. Alternatively, the flux measurement in \cite{2018ApJ...854...99W} may be inaccurate, provided that LSXPS only gives an upper limit. The difference may also be explained by the substantial absorption and the contamination of the ACIS detector at low energies and/or the fact that the source lies near the chip gap. No optical-IR counterpart is found within the $1.2\arcsec$ PU of the X-ray source.
Without any matching counterpart, the X-ray source is classified as an NS with CT=1.7.

\subsubsection{4FGL J1736.1--3422}

4FGL J1736.1--3422 
has been associated with the $\gamma$-ray pulsar PSR J1736--3422, identified by the 3PC \citep{2023arXiv230711132S}. No X-ray source is seen in the ACIS-I image within  $10''$ from the position of the $\gamma$-ray pulsar.

Only one significant X-ray source, CXO J173609.9--342419 (D7), is detected  (with S/N=3.8)  within the 95\% PU of J1736.1--3422  (see panel 4 in Figure~\ref{fig:FGL-fields-zoomin}). This source has an association (D7-1) with an optical source, Gaia DR3 4053560004787094528, but with a fairly low association probability, $P_{\rm i}=0.45$, given the X-ray PU of $1.4\arcsec$ and the high density of Gaia sources in this field.
D7 is classified as an LMXB without the optical counterpart, albeit with a  very low confidence (CT=0.7). With the optical counterpart,  
 D7-1 is very confidently classified as a CV (with CT=10.4).

Running \texttt{srcflux}  at the position of PSR J1736--3422  yields an upper limit (for the observed/absorbed flux) of $6.3\times10^{-15}$\,erg\,cm$^{-2}$\,s$^{-1}$ at the 95\% confidence level.

\subsubsection{4FGL J1639.3--5146}

The Giant Metrewave Radio Telescope (GMRT)  study by \cite{2016MNRAS.461.1062F} revealed 
a bright ($174.9\pm16.3$\,mJy at 150\,MHz) and compact radio source, 
TGSSADR J163923.8--514634 (with R.A.=16:39:23.83(0.03) and decl.=--51:46:34.1(2.8), which is also listed in the GMRT all-sky  survey TGSS ADR; \citealt{Intema2017}), within the LAT source PU. 
A potential optical counterpart with a B-band magnitude of 18.8,  cataloged in the USNO-B1.0 survey \citep{Monet2003},  is located 2.1$\arcsec$ from the TGSSADR source.
The radio source 
landed within the chip gap.  
No  statistically significant  X-ray sources are detected in the ACIS-I image within the 95\% PU of 4FGL J1639.3--5146.
We note that an X-ray source, CXO J163932.2--514620 (E8), was detected outside the PU of 4FGL J1639.3--5146, but inside the PU of 3FGL J1639.4--5146. This X-ray source has no associations at lower frequencies. 
It is classified as an LMXB
(with CT=2.3).

\subsubsection{4FGL J1203.9--6242}

4FGL J1203.9--6242 has been associated with the $\gamma$-ray pulsar PSR J1203--6242, discovered by the 3PC \citep{2023arXiv230711132S}. The pulsar is not significantly detected by \texttt{wavdetect} at the preliminary localization from the 4FGL-DR4 catalog within an assumed PU of $10''$, with only a slight hint of enhanced emission seen in the ACIS-I image.  Running \texttt{srcflux}  at the position of PSR J1203--6242  yields an upper limit for the flux of $10^{-14}$\,erg\,cm$^{-2}$\,s$^{-1}$ at the 95\% confidence level within the 90\% encircled PSF aperture.

 Only one X-ray source, CXO J120355.5--624231 (F2), is significantly (S/N$=5.1$) detected within the 95\% $\gamma$-ray PU (see  panel 5 in Figure~\ref{fig:FGL-fields-zoomin}). 
 Because F2 is only $\sim13\arcsec$ away from PSR J1203--6242,
 it might be associated with PSR J1203--6242 if the (unpublished) uncertainty of the pulsar position is large enough. 
 This X-ray source has a potential association (F2-1) with an optical source,  
 Gaia DR3 6057424069589180032.
Without the optical association, MUWCLASS classifies F2 as an NS with low confidence  (CT=1.3). With the association,   F2-1 is classified as a CV with high confidence (CT=8.5). Given that the positional offset of the Gaia source  (0.12$''$) is much smaller than the PU of the X-ray source (0.8$''$), we estimate the association probability, $P_{\rm i}$, to be high (81\%), and hence, the latter classification is more likely. A CV is unlikely to contribute to the $\gamma$-ray emission from this region.

\subsubsection{4FGL J0854.8--4504}

Using an ML approach, 3FGL J0854.8--4503 (which later became 4FGL J0854.8--4504) has been classified as 
a young pulsar based solely on its $\gamma$-ray properties 
\citep{2016ApJ...820....8S}. However, no $\gamma$-ray or radio pulsations were  reported from this region.  
LSXPS detected three marginal X-ray sources, LSXPS J085451.1--450418, LSXPS J085448.8--450524, and LSXPS J085456.8--450348, within the $\gamma$-ray PU.%, with only poor detection flags. 

Only one X-ray source, CXO J085457.4--450350 (G1), is significantly detected (S/N$=$3.8) in the ACIS-I image  within the 95\% PU of 4FGL J0854.8--4504. The X-ray source also coincides with LSXPS J085456.8--450348 with the observed fluxes  $2.6\pm1.2\times10^{-14}$\,erg\,cm$^{-2}$\,s$^{-1}$ (from CXO) and  $3.7^{+1.8}_{-1.7}\times10^{-13}$\,erg\,cm$^{-2}$\,s$^{-1}$ (from Swift XRT), 
suggesting that the source is variable. It is located near the edge of the ellipse and has an association (G1-1) with an optical (and IR) counterpart, Gaia DR3 5331365758050152704 (2MASS 08545736-4503504 and CatWISE2020 J085457.36-450350.4). Without the optical--IR association, G1 is classified as an NS with fairly high confidence (CT=4.8). However, when the optical--IR association is included, G1-1 is classified as an LM-STAR with  CT=2.3.  Given that the positional offset of the Gaia source  (0.07$''$) is much smaller than the PU of the X-ray source (0.97$''$), we estimate the association probability to be high (96\%), and hence, the LM-STAR classification is 
more likely. However, at the distance of $2100\pm100$ pc \citep{2021AJ....161..147B}, the X-ray luminosity would be in the $(1-30)\times10^{31}$ ergs s$^{-1}$ range (corresponding to the CXO and Swift flux measurements), which is unusually high for a solitary LM-STAR outside a major coronal flare. We do not see any evidence of such a flare during the CXO observation. LSXPS does not detect a significant variability in this source either. Moreover, given that the stellar spectral type is $\sim$A (this is within our LM-STAR class definition) according to Gaia DR3, significant coronal activity is not expected (see, however, \citealt{2023ApJ...948...59C}, who reported several X-ray bright A stars).

We note that 4FGL J0854.8--4504 is located within the Vela SNR extent ($r\approx4^\circ$), and it could be that some part of the SNR shell is detected by LAT.

\subsubsection{4FGL J2035.0+3632}

4FGL J2035.0+3632 has been associated with the $\gamma$-ray pulsar PSR J2034+3632,  discovered by the 3PC \citep{2023arXiv230711132S}. This is an MSP with a rather low spin-down power $\dot{E}\approx10^{33}$\,erg\,s$^{-1}$. Hence, its nondetection in X-rays is not surprising, especially, if the distance to the pulsar is substantial (it cannot be reliably determined from $\gamma$-rays alone).   No  X-ray sources were detected by \texttt{wavdetect} above the $3\sigma$ significance threshold within the 95\% PU ellipse of 4FGL J2035.0+3632. 
Running \texttt{srcflux}  at the position of PSR J2034+3632  yields an upper limit for the observed (absorbed) flux of $3.9\times10^{-15}$\,erg\,cm$^{-2}$\,s$^{-1}$ at the 95\% confidence level.

\subsubsection{4FGL J2041.1+4736}

3FGL J2041.1+4736 (which later became 4FGL J2041.1+4736) has been classified as a young pulsar based solely  on its $\gamma$-ray properties, 
using an ML approach  
\citep{2016ApJ...820....8S}.

Only one X-ray source, CXO J204116.7+473658 (I2),  is significantly detected ($\approx6\sigma$) in the ACIS-I image  within the 95\% PU ellipse of 4FGL J2041.1+4736 (see  panel 7 in Figure~\ref{fig:FGL-fields-zoomin}). 
The X-ray source has an association (I2-1) with an IR counterpart, CatWISE2020 J204116.56+473659.8, with a relatively high $P_{\rm i}=0.82$. Without the association, I2 is classified as an NS with low confidence (CT=1.1).  If the IR association is included, the classification changes to a confident  AGN (CT=8.8). 
The detection of a radio source, MSC J204116.66+473659.2, with a flux of $0.65\pm0.12$\,mJy at 5.9 GHz  and  spectral index $\alpha=0.35\pm0.25$ \citep{2023ApJ...943...51B}, coincident with the X-ray source, 
supports the AGN classification for CXO J204116.7+473658. The identification of the only X-ray counterpart as an AGN suggests that the 4FGL source may be an AGN.

\subsubsection{4FGL J1317.5--6316}

No X-ray sources is detected in the ACIS-I image for 4FGL J1317.5--6316 above the \texttt{wavdetect} $3\sigma$ threshold within its 95\% PU ellipse.

\subsubsection{4FGL J1329.9--6108}

3FGL J1329.8--6109 (which later became 4FGL J1329.9--6108) was classified as an MSP using an ML-based classification based only on its $\gamma$-ray properties \citep{2016ApJ...820....8S}. The source has been observed by Swift XRT, and was associated with SwF3 J132939.6--610735, detected within the 95\% PU of 3FGL J1329.8--6109 \citep{2019ApJ...887...18K,2021AJ....161..154K}, but outside the 95\% PU of 4FGL J1329.9--6108. 
SwF3 J132939.6--610735 was classified as a 
pulsar with a probability of 0.94 based on its X-ray and $\gamma$-ray properties using an ML approach \citep{2019ApJ...887...18K}. 
However, the detection of SwF3 J132939.6--610735 is likely severely affected by 
optical loading\footnote{Optical loading happens when significant charge is caused in the CCD by enough optical photons from bright optical sources, affecting the onboard detection of X-rays. For a more technical description, see \url{https://www.swift.ac.uk/analysis/xrt/optical_loading.php}.} (\citealt{2021AJ....161..154K,2021RNAAS...5..259H}) 
from the nearby (offset by 10\arcsec), bright  (G=9 mag) optical star, Gaia DR3 5868669568445358336,   given  the very soft X-ray spectrum of the XRT source with a high flux. The origin of the 10$\arcsec$ between the star and SwF3 J132939.6--610735 is unclear because the coordinates and the uncertainties of the XRT source position are not listed in \cite{2019ApJ...887...18K}.

We note that LSXPS does not detect SwF3 J132939.6--610735, but does detect an X-ray source, LSXPS J132938.8--610745, which is $\approx10''$ offset from SwF3 J132939.6--610735. The origin of this offset is unclear, but LSXPS J132938.8--610745 lies only $1\farcs8$ offset from the bright star, suggesting that its X-ray properties are impacted by optical loading. LSXPS J132938.8--610745 has an observed (absorbed) flux of $1.2^{+0.3}_{-0.2}\times10^{-13}$\,erg\,cm$^{-2}$\,s$^{-1}$ in the 0.3--10\,keV energy band. It is coincident with  one of the X-ray sources, CXO J132939.1--610744 (K4), detected in our \chandra\ observations, which has a consistent flux of $\sim1.5\times10^{-13}$\,erg\,cm$^{-2}$\,s$^{-1}$ in the 0.5--7\,keV energy band. The positions of both of these X-ray sources are compatible with the bright star. Because the ACIS-I spectrum should not be affected by the optical loading\footnote{The ACIS detector has a filter that blocks optical light. Moreover, a contaminant has been accumulating on this filter that further reduces any impact of the optical light. (see \url{ https://cxc.cfa.harvard.edu/ciao/why/acisqecontamN0014.html}.)}, we conclude that X-rays must be coming from the star and likely have a coronal origin,  given the soft spectrum (see Table \ref{tab:X-ray-Class}) and the relatively low luminosity of $1.3\times10^{29}$ erg s$^{-1}$ at the Gaia-inferred distance of 84 pc \citep{2021AJ....161..147B}.

No X-ray sources were detected by \texttt{wavdetect}  above the 
$3\sigma$ threshold in the ACIS-I image 
within the 95\% error ellipse of 4FGL J1329.9--6108. 
Six X-ray sources were detected within the PU of 3FGL J1329.8--6109, but outside the PU of 4FGL J1329.9--6108, which we discuss in Section\,\ref{sec:other-sources} and Table\,\ref{tab:X-ray-Class}.

\subsubsection{4FGL J0744.0--2525}

4FGL J0744.0--2525 
has been associated with the $\gamma$-ray pulsar PSR J0744--2525, discovered in LAT data by the 3PC \citep{2023arXiv230711132S}.  This pulsar is old, with a characteristic age ($\tau_{\rm c}$) of about 1.6\,Myr, and it has a fairly low $\dot{E}=4.7\times10^{34}$\,erg\,s$^{-1}$.  No X-ray sources were detected by \texttt{wavdetect} above the $3\sigma$ significance threshold within the 95\% PU ellipse of 4FGL J0744.0--2525. Because the pulsar is old, its nondetection in X-rays is not surprising. 
There are no photons in the 0.5--7\,keV energy band within the 3$\arcsec$ radius circle around the position of PSR J0744--2525. We have used a circular annulus   with  a radius of 30$\arcsec$ centered on PSR J0744--2525 to measure the local background and obtained a mean background surface brightness of $0.005\pm0.001$\,counts\,arcsec$^{-2}$. 
 Thus, within the r = 1$''$ extraction aperture (corresponding to $\approx$90\% encircled PSF at 1.5 keV), the background is very small 
 ($\approx$0.016 counts). 
 For a Poissonian distribution, the
nondetection with zero counts translates into an upper limit
$N < - \ln(1-CL)$ counts at the confidence level CL, which
gives the 95\% count rate upper limit of $3\times10^{-4}$ counts s$^{-1}$. For an absorbed PL with $\Gamma=2$ and N$_H=10^{22}$ cm$^{-2}$, this corresponds to an absorbed flux of  $6\times10^{-15}$ erg s$^{-1}$  cm$^{-2}$.

\subsubsection{4FGL J1358.3--6026}

4FGL J1358.3--6026  has been associated with the $\gamma$-ray pulsar PSR J1358--6025, discovered in the LAT data by the 3PC \citep{2023arXiv230711132S}. Although \texttt{wavdetect} does not detect the pulsar counterpart in the ACIS-I image above the $3\sigma$ threshold, extended X-ray emission is seen in the ACIS-I image, with its brightest part covering the $\gamma$-ray  pulsar position. We find that this brightest part  is detected by \texttt{wavdetect} as CXO J135825.9--602555 (M8)  with a lower significance of $2.7\sigma$. We note that the \texttt{wavdetect} search is more sensitive for point sources than for extended sources. We describe the properties of this extended source in more detail below. CXO J135825.9--602555 is classified as an NS with  CT$=$2.2, and it lacks any lower-frequency counterparts.

Two other X-ray sources, CXO J135834.6--602800 (M4) and CXO J135817.2--602504 (M6), have been detected by \texttt{wavdetect} above the $3\sigma$ significance threshold in the ACIS-I image within the 95\% PU ellipse of the 4FGL source (see panel 8 in Figure~\ref{fig:FGL-fields-zoomin}). Both sources are situated near the periphery of the $\gamma$-ray error ellipse and are tentatively classified as AGNs without any matching counterparts, albeit with very low confidence (CT=0.3 and CT=0.2).

\subsection{Extended Sources}
\label{sec:extended}

\begin{figure}
\includegraphics[width=0.99\columnwidth]{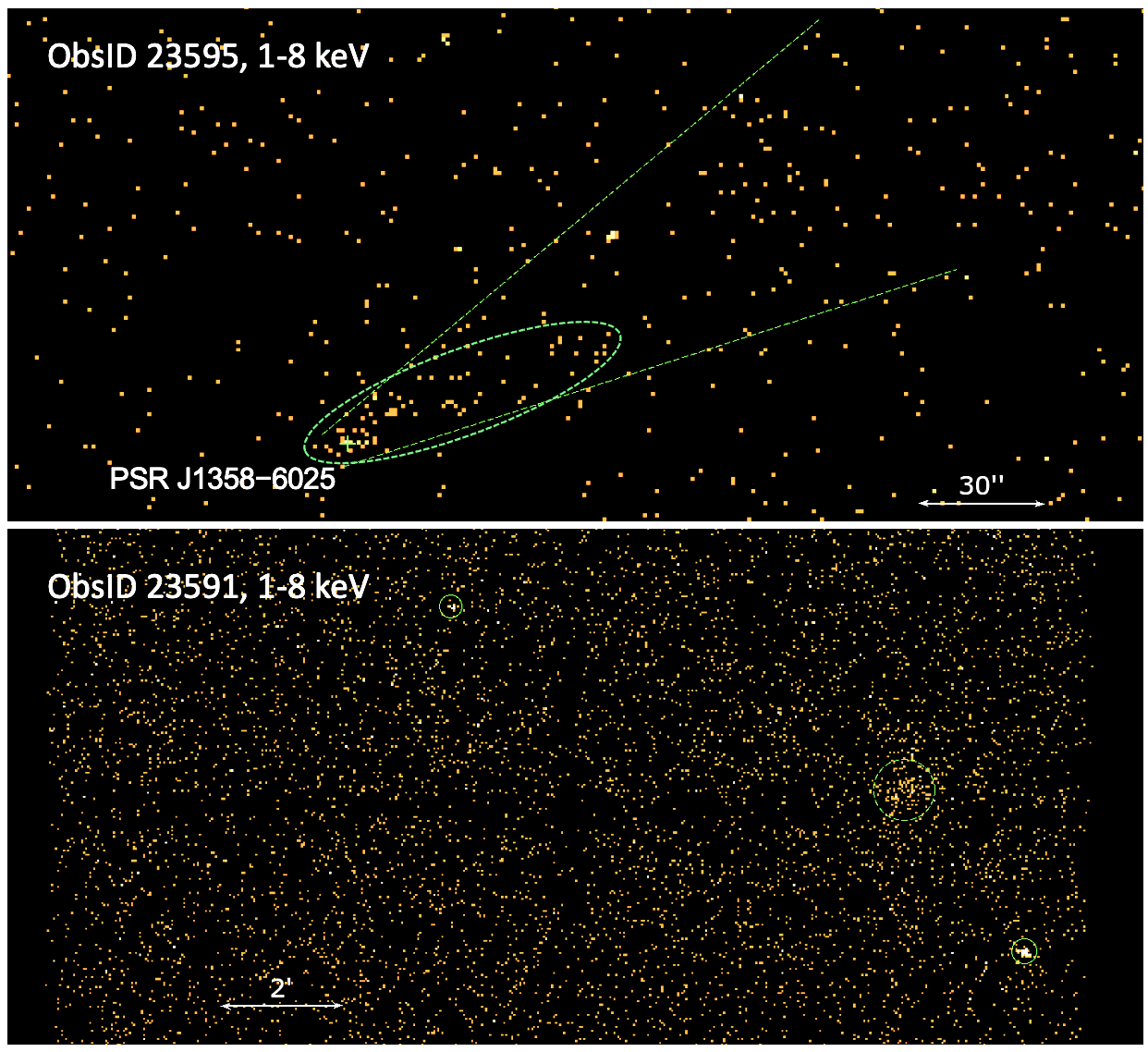}
\caption{Extended sources  in the ACIS-I images from ObsIDs 23595 (top) and ObsIDs  23591 (bottom). The ellipse and circle show  spectral extraction regions (see text). In the top panel, the cross marks the position of the  pulsar (see Table \ref{tab:4FGLDR3-NSs}).  } 
\label{fig:extended}
\end{figure}

Figure \ref{fig:extended} shows two extended sources in the ACIS-I images from ObsIDs 23591 (4FGL J2041.1+4736) and 23595 (4FGL J1358.3--6026). 

One of the extended sources (identified above as CXO J135825.9--602555) is clearly associated with the 300\,kyr old pulsar PSR J1358--6025. The X-ray emission near the pulsar location is brightest, but it does not look like a point source, suggesting that the PWN emission is predominantly detected, although on a larger scale, the emission is quite faint, and its overall morphology (emission is confined between the two dashed lines in the top panel of Figure \ref{fig:extended}) suggests that it may be a tail that is confined within the bow-shock PWN associated with the supersonically moving pulsar \citep{Kargaltsev_Pavlov2008}. The spectrum extracted from the brightest part of the PWN  can be fit by an absorbed PL model with a photon index $\Gamma=2.4\pm0.8$ and $n_{H}=2.3_{-1.1}^{+1.5}\times10^{22}$ cm$^{-2}$. The unabsorbed flux is $2.8_{-1.6}^{+5.3}\times10^{-13}$ erg cm$^{-2}$ s$^{-1}$ in 0.5--8 keV. Because the distance to the pulsar is not constrained, the luminosity is difficult to estimate accurately. Given its location in the Galactic plane ($l=311.1$, $b=1.4$), we use a fiducial distance of $8$ kpc to estimate the X-ray luminosity $L_{\rm X}\approx2\times10^{33}$\,erg\,s$^{-1}$ and the X-ray PWN radiative efficiency $\approx4\times10^{-3}$. These parameters are fairly typical for a PWN \citep{Kargaltsev_Pavlov2008}.  

The other extended source has a more isotropic appearance with a gradually decreasing surface brightness, starting from the center of the source. The source spectrum (extracted from the $r=0.5'$ circular aperture shown in Figure~\ref{fig:extended}) is hard and can be described either by an absorbed PL with $\Gamma=0.4_{-0.3}^{+0.5}$ and  $n_{H}=0.2_{-0.2}^{+0.9}\times10^{22}$ cm$^{-2}$ or by thermal plasma emission (e.g., {\tt mekal} in XSPEC) with kT$>10$ keV and $n_{H}=(2\pm1)\times10^{22}$ cm$^{-2}$. The unabsorbed flux is $3.2_{-0.4}^{+0.6}\times10^{-13}$ erg cm$^{-2}$ s$^{-1}$ in 0.5--8 keV. Given the hard spectrum and the appearance in the X-ray image, we conclude that the source likely is a cluster of galaxies projected near the Galactic plane. It is therefore unlikely to be the source of GeV $\gamma$-rays by itself. The two X-ray sources seen in its vicinity are AGNs (based on our classifications and supported by their radio counterparts in the TGSS GMRT radio survey) and could produce some GeV emission. %However, they are both located outside the 4FGL source PU.     

\subsection{ X-Ray Sources outside the $\gamma$-ray Error Ellipses}
\label{sec:other-sources}

Eighty-four X-ray sources were detected by \texttt{wavdetect} above 
the $3\sigma $ significance threshold
in the ACIS-I images outside the 4FGL-DR4 PUs. 
After we removed 24 sources with X-ray PU larger than 3\arcsec, the 60 remaining X-ray sources  
were matched to 116 associations (including the cases without a counterpart). Below, we briefly discuss 
the classifications of the 60 X-ray sources with confident classifications shown in Table~\ref{tab:X-ray-Class}.

For 20 X-ray sources, no counterparts were found within their X-ray PUs. 
Three  of them are confidently classified as 
NSs (CXO J190706.3+072001, B2, CT=3.8; CXO J132935.6--611106, K16, CT=3.1; and CXO J074428.2–252744, L2, CT=2.3), while the other two are confident LMXBs (E8 and CXO J204112.7+473012, I5, CT=4.4). Another 20 X-ray sources are confidently classified as 14 NSs and 6 LMXBs when no counterparts are assumed, while the classifications change to 6 AGNs, 5 LM-STARS, 3 CVs, and 6 nonconfident classifications if the counterparts are taken into account. The association probabilities for the
 confident classifications are given in Table~\ref{tab:X-ray-Class}, if available. For the 
remaining 20 X-ray sources, there are no confident classifications without counterparts.   
However, all of these sources have possible counterparts with lower-frequency sources. 
When the counterparts are taken into account, 11 of them are confidently classified as 4 LM-STARs, 3 AGNs, 2 CVs, and 2 YSOs.
Three further sources (CXO J163906.8--515230 (E5), CXO J120419.8--623718 (F7), and CXO J132924.2--610417 (K12)) have more than one confident classification based on their multiple possible associations (see Table \ref{tab:X-ray-Class} for details). 

The sources are confidently classified as either NSs or LMXBs mostly due to the lack of their counterparts and/or because they are very faint. This is likely due to a bias caused by the attempt to classify sources that are systematically fainter than the sources in our TD. These fainter sources are underrepresented  in the TD because it is more difficult to unambiguously classify them, and hence, they are not included in the TD. 
It is not surprising that a large fraction of faint sources does not have MW counterparts simply because of the insufficient sensitivity of the optical and IR all-sky surveys, which are currently used, exacerbated by the significant extinction in the Galactic plane. The MUWCLASS
pipeline 
tends to interpret the lack of MW counterparts as a sign of an NS (because NSs lack any detectable counterpart in the optical, near-IR, and IR surveys). Therefore, the NS classifications of 25 sources lacking counterparts should be taken with caution.   They include one relatively bright source, CXO J133035.1--611121 (K2), which is more likely to be an NS than others because its spectrum shows evidence of nonthermal emission and low absorption (see details in Table~\ref{tab:X-ray-Class-bright}). CXO J133035.1--611121 is not only confidently classified as an NS without a counterpart, it is also classified as an NS even with a possible Gaia counterpart (Gmag$=$20.4, $P_{\rm i}=78\%$) with CT$=1.7$. The other relatively bright sources are likely AGN/YSO based on their large absorbing column densities and MW counterparts (see Table~\ref{tab:X-ray-Class-bright}). The optical counterpart of CXO J163910.2-514555 (E2-1) has a Gaia parallax ($\omega=0.09\pm0.2$) consistent with 0, suggesting that it may be an AGN that is misclassified as a CV.

By individually classifying multiple associations to the same X-ray source,  
we have confidently identified 10 LM-STARs and 10 AGNs. Of these, 9 LM-STARs have accurate distance measurements  (ranging from 84\,pc to 2.1\,kpc) in the Gaia eDR3 catalog \citep{Bailer-Jones2020}.
The corresponding $L_{\rm X}$ range from 1.2$\times10^{29}$\,erg\,s$^{-1}$ to 3.0$\times10^{31}$\,erg\,s$^{-1}$, which is typical for LM-STARs \citep{2016ApJ...830...44N}. Of the 10 AGNs, none has accurate distance measurements in Gaia DR3, consistent with their extragalactic nature. We also cross-correlated all X-ray sources to radio catalogs and find that CXO J204023.6+473315 (I1) matches an extended radio source that is reported in multiple radio catalogs, including the Very Large Array Sky Survey Quicklook catalog \citep{2021ApJS..255...30G} and the SPECFIND V3.0 catalog \citep{2021A&A...655A..17S}. The source shows a flat spectrum between 73 and 150 MHz, but a steep/broken spectrum above 232 MHz (up to 4.85 GHz), as shown in its radio spectral energy distribution from \cite{2021A&A...655A..17S}, which is consistent with a radio AGN scenario. The other source, CXO J204119.1+473854 (I3), also matches an unresolved radio source from \cite{2021A&A...655A..17S}, showing a steep spectrum from 150 MHz up to 4.85 GHz. I3 is classified as an NS with low confidence and lacks a matching counterpart. It is therefore not shown in Table\,\ref{tab:X-ray-Class}. 

None of the sources classified as CVs has a good distance measurement.  
The LMXB candidate, F7-2, associated with CXO J120419.8--623718, and one YSO candidate, M1-1, associated with CXO J135857.8--602736, have good distance measurements of $2076^{+660}_{-450}$\,pc and $2311^{+254}_{-181}$\,pc, respectively, corresponding to $L_{\rm X}$ of $1.1\times10^{31}$ and $1.2\times10^{32}$\,erg\,s$^{-1}$.

\section{Summary and Conclusions}
\label{sec:concl}

\begin{itemize}
    
    \item Eight of the 4FGL sources contain $\gamma$-ray pulsars within their position uncertainty ellipses,  three of which are  detected in X-rays, and one shows a hint of X-ray emission. 
    \item We have detected two extended X-ray sources, for one of which (associated with 4FGL J1358.3--6026/PSR J1358--6025), the  properties are consistent with a PWN of a bow-shock type. 
    The other extended X-ray source is likely to be a galaxy cluster located outside the  4FGL J2041.1+4736 PU. 
    \item We detected 93 X-ray sources above the $3\sigma$ significance threshold in the  CXO ACIS-I observations of 13 GeV $\gamma$-ray sources. Nine of the CXO sources are  located within the error ellipses of seven  4FGL sources. Of these  seven, only  two 4FGL sources (J0854.8--4504, and  J2041.1+4736) lack a $\gamma$-ray pulsar association. In each of these, only one X-ray source is confidently detected with the 4FGL 95\% PU ellipse (G1 and I2). G1 is confidently classified  as an NS without the counterpart. With the counterpart (the more likely case), it is a stellar type source, but given the surprisingly large luminosity, it may be a binary. An optical spectroscopic observation might be useful to explore the putative binary nature, and alternatively, a deeper X-ray and/or radio observation can be used to search for pulsations from the possible NS scenario that might be associated with the GeV emission. 
    I2 is likely to be an AGN given its MW classification and the presence of the radio counterpart.      
    \item For X-ray sources detected outside of the LAT source PUs, we provide classifications, some of which are consistent with isolated or binary compact objects, depending on the confidence with which they are associated with sources at lower frequencies. 

\end{itemize}

 We have selected a sample of bright $\gamma$-ray sources (i.e., only 17$\%$ of GeV sources in the 4FGL-DR4 catalog have $G_{100}>10^{-11}$ c.g.s.) 
for this study. Thus, it is not surprising that blind pulsation searches have been successful in establishing the pulsar nature for most of them. However, for fainter  $\gamma$-ray sources,   blind pulsation searches are progressively more difficult, and  identifying a counterpart at lower energies becomes crucial. We demonstrate that the ML approach can be used to identify promising X-ray candidates in some cases at least. Our analysis also shows that partly due to the degrading sensitivity of ACIS caused by the accumulating contamination, 10\,ks observations can no longer provide enough photons for reliable classifications or  even for significant detections,  especially if the exact location of the source is not known a priori. On the other hand, the very low ACIS background and subarcsecond resolution of CXO can resolve the faint extended emission, which facilitates the identifications. Finally, for three pulsars detected in $\gamma$-rays, we were able to estimate the X-ray fluxes, which can be used to explore the connection between the $\gamma$-ray and X-ray emission in pulsars.

\begin{acknowledgments}

This work was supported by the National Aeronautics and Space Administration through Chandra Awards GO8-19058X, GO1-22072A, GO1-22072B, and AR0-21007X  issued by the Chandra X-ray Center, operated by the Smithsonian Astrophysical Observatory for the National Aeronautics Space Administration. J.H. acknowledges support from NASA under award number 80GSFC21M0002.

\end{acknowledgments}

\begin{deluxetable*}{llcccccccccccc}
\tabletypesize{\scriptsize}
\tablecaption{Summary of the MW Properties and Classifications of CXO Sources within the 4FGL  95\% PU Ellipses 
\label{tab:X-ray-Class-errorellipse}}
\tablehead{
\colhead{CXO name} & \colhead{A-name}  & \colhead{S/N$^{\rm a}$} & \colhead{PU$^{\rm b}$} & \colhead{$F_{\rm b}^{\rm c}$} & \colhead{HR$_{\rm hms}^{\rm d}$}  & \colhead{G$^{\rm e}$} & \colhead{J$^{\rm f}$}  & \colhead{W1$^{\rm g}$}   &    \colhead{d$^{\rm h}$}  & \colhead{$P_{\rm i}^{\rm i}$} & \colhead{$P_{\rm c}^{\rm i}$}  & \colhead{Class$^{\rm j}$} & \colhead{CT}  \\
& &  & (arcsec) & ($10^{-14}$\,cgs)  &  & (mag) & (mag) & (mag) & (pc) &  &  & & }
\startdata
J130621.8--604328$^\dag$ &  A9 &             3.0 &    1.15 &   1.19 &     -0.25 &     \nodata &     \nodata &      \nodata &  \nodata &    \nodata &         \nodata &      NS &   1.1 \\
J130621.8--604328$^\dag$ & A9-1 &             3.0 &    1.15 &   1.19 &     -0.25 &   16.97 &   14.43 &    13.07 & $2002_{-297}^{+670}$ &   0.83G &        0.17G &     YSO &   2.2 \\
 J110500.5--603715 &   C6 &             1.7 &    1.23 &   0.25 &     -0.99 &     \nodata &     \nodata &      \nodata &  \nodata &    1 &         0 &      NS &   1.7 \\
 J173609.9--342419 &   D7 &             3.8 &    1.44 &   4.14 &      0.87 &     \nodata &     \nodata &      \nodata &  \nodata &    \nodata &         \nodata &    LMXB &   0.7 \\
 J173609.9--342419 & D7-1 &             3.8 &    1.44 &   4.14 &      0.87 &   18.93 &     \nodata &      \nodata & $1247_{-366}^{+528}$ &   0.45G &        0.55G &      CV &  10.4 \\
 J120355.5--624231 &   F2 &             5.1 &    0.83 &   5.45 &      0.32 &     \nodata &     \nodata &      \nodata &   \nodata &    \nodata &         \nodata &      NS &   1.3 \\
 J120355.5--624231 & F2-1 &             5.1 &    0.83 &   5.45 &      0.32 &   18.59 &     \nodata &      \nodata &   \nodata &   0.81G &        0.19G &      CV &   8.5 \\
J085457.4--450350$^\dag$ &   G1 &             3.8 &    0.97 &   2.62 &      0.43 &     \nodata &     \nodata &      \nodata &   \nodata &    \nodata &         \nodata &      NS &   4.8 \\
J085457.4--450350$^\dag$ & G1-1 &             3.8 &    0.97 &   2.62 &      0.43 &   15.60 &   12.94 &    12.06 & $2103_{-107}^{+109}$ &   0.96G &        0.04G & LM-STAR &   2.3 \\
J204116.7+473658$^\dag$ &   I2 &             6.0 &    0.81 &   8.46 &      0.83 &     \nodata &     \nodata &      \nodata &  \nodata &    \nodata &         \nodata &      NS &   1.1 \\
J204116.7+473658$^\dag$ & I2-1 &             6.0 &    0.81 &   8.46 &      0.83 &     \nodata &     \nodata &    15.57 &  \nodata &   0.82C &        0.18C &     AGN &   8.8 \\
 J135834.6--602800 &   M4 &             4.1 &    0.74 &   3.14 &      0.84 &     \nodata &     \nodata &      \nodata &  \nodata &    1 &         0 &     AGN &   0.3 \\
 J135817.2--602504 &   M6 &             3.1 &    0.87 &   2.71 &      0.80 &     \nodata &     \nodata &      \nodata &  \nodata &    1 &         0 &     AGN &   0.2 \\
 J135825.9--602555 &   M8 &             2.7 &    0.76 &   1.48 &      0.90 &     \nodata &     \nodata &      \nodata &  \nodata  &    1 &         0 &      NS &   2.2 \\
\enddata
\tablecomments{$^{\rm a}$ The X-ray detection significance from \texttt{wavdetect}. $^{\rm b}$ The X-ray positional uncertainty of the 95\% confidence level. $^{\rm c}$ The observed broadband X-ray flux in the 0.5--7\,keV energy band. $^{\rm d}$ Hardness ratio ($F_{\rm h}-F_{\rm m}-F_{\rm s}$)/($F_{\rm h}+F_{\rm m}+F_{\rm s}$), where $F_{\rm h}$, $F_{\rm m}$, and $F_{\rm s}$ are the energy fluxes in the hard (2.0--7.0\,keV), medium (1.2--2.0\,keV), and soft (0.5--1.2\,keV) energy bands. $^{\rm e}$ Gaia DR3 G-band magnitude. $^{\rm f}$ 2MASS J-band magnitude. $^{\rm g}$ WISE W1-band magnitude. $^{\rm h}$ The distances for sources with Gaia DR3 associations are taken from \cite{2021AJ....161..147B} and are only listed when the ratio of the parallax to its standard error is $\ge3$; 
$^{\rm i}$ $P_{\rm c}$ is the chance coincidence probability, and $P_{\rm i}$ is the association probability calculated for a specific catalog as indicated by a letter (G for Gaia, C for CatWISE2020, T for 2MASS, and A for AllWISE).
$^{\rm j}$ The predicted classification. The column definitions are the same for the same columns in Tables \ref{tab:X-ray-Class}, \ref{tab:4FGLDR3-NSs}, and \ref{tab:X-ray-Class-bright}. 
All X-ray sources listed in this table are not significantly variable (the variability probability of each X-ray observation calculated from the Kuiper test is $<95\%$).
X-ray sources whose names are marked by $^\dag$ are located within the PU of 4FGL sources currently lacking $\gamma$-ray pulsar associations (see Figure \ref{fig:FGL-fields-zoomin}).   
}
\end{deluxetable*}

\begin{deluxetable*}{llcccccccccccc}
\tabletypesize{\scriptsize}
\tablecaption{Summary of the MW Properties and Confident Classifications of X-Ray Sources outside the Fermi Error Ellipses with X-Ray PU$\le3\arcsec$
\label{tab:X-ray-Class}}
\tablehead{
\colhead{CXO name} & \colhead{A-name}  & \colhead{S/N} & \colhead{PU} & \colhead{$F_{\rm b}$} & \colhead{HR$_{\rm hms}$}  & \colhead{G} & \colhead{J}  & \colhead{W1}   &    \colhead{d}  & \colhead{$P_{\rm i}$} & \colhead{$P_{\rm c}$}  & \colhead{Class} & \colhead{CT}  \\
& &  & (arcsec) & ($10^{-14}$\,cgs)  &  & (mag) & (mag) & (mag) & (pc) &  &  & & }
\startdata
 J130657.3--604823 &    A1 &             5.5 &    1.56 &   6.12 &      0.62 &     \nodata &     \nodata &      \nodata &  \nodata &   \nodata &   \nodata &      NS &   3.3 \\
 J130657.3--604823 &  A1-1 &             5.5 &    1.56 &   6.12 &      0.62 &   19.00 &     \nodata &    12.57 &  \nodata & 0.60G & 0.40G &     AGN &   3.0 \\
 J130703.7--604653 &    A3 &             5.3 &    1.48 &   4.82 &      0.72 &     \nodata &     \nodata &      \nodata &  \nodata &   \nodata &   \nodata &      NS &   2.6 \\
 J130703.7--604653 &  A3-1 &             5.3 &    1.48 &   4.82 &      0.72 &     \nodata &     \nodata &    13.80 &  \nodata & 0.85C & 0.15C &     AGN &  27.7 \\
 J130550.6--604257 &    A4 &             3.9 &    1.69 &   3.62 &      0.26 &     \nodata &     \nodata &      \nodata &  \nodata &   \nodata &   \nodata &    LMXB &   2.8 \\
 J130642.2--604910 &    A5 &             3.8 &    1.90 &   2.94 &      0.83 &     \nodata &     \nodata &      \nodata &  \nodata &   \nodata &   \nodata &      NS &   2.1 \\
 J130642.2--604910 &  A5-1 &             3.8 &    1.90 &   2.94 &      0.83 &   18.89 &     \nodata &      \nodata &  \nodata & 0.54G & 0.46G &      CV &   6.9 \\
 J190712.8+072414 &    B1 &             5.7 &    2.80 &   6.56 &      0.73 &     \nodata &     \nodata &      \nodata &  \nodata &   \nodata &   \nodata &      NS &   2.7 \\
 J190706.3+072001 &    B2 &             4.6 &    1.85 &   6.70 &      0.92 &     \nodata &     \nodata &      \nodata &  \nodata &     1 &     0 &      NS &   3.8 \\
 J190658.2+072330 &    B3 &             4.6 &    1.57 &   3.07 &     -0.34 &     \nodata &     \nodata &      \nodata &  \nodata &   \nodata &   \nodata &    LMXB &   8.1 \\
 J190658.2+072330 &  B3-2 &             4.6 &    1.57 &   3.07 &     -0.34 &   17.44 &   12.86 &    11.66 &  \nodata & 0.44G & 0.07G & LM-STAR &   2.0 \\
 J173539.2--342513 &  D2-1 &             5.9 &    1.85 &   8.94 &      0.66 &   16.47 &   11.88 &      \nodata &  \nodata & 0.29G & 0.52G &     YSO &   2.4 \\
 J173550.3--341840 &  D3-1 &             5.0 &    1.81 &   5.46 &      0.90 &     \nodata &     \nodata &    13.17 &  \nodata & 0.79C & 0.21C &     AGN &  23.7 \\
 J173624.2--342050 &    D6 &             3.9 &    1.78 &   4.00 &      0.91 &     \nodata &     \nodata &      \nodata &  \nodata &   \nodata &   \nodata &      NS &   3.6 \\
 J163830.6--514243 &    E1 &             9.0 &    2.13 &  28.34 &      1.00 &     \nodata &     \nodata &      \nodata &  \nodata &   \nodata &   \nodata &      NS &   4.7 \\
 J163830.6--514243 &  E1-2 &             9.0 &    2.13 &  28.34 &      1.00 &     \nodata &     \nodata &    12.73 &  \nodata & 0.92A & 0.08A &     AGN &   9.7 \\
 J163910.2--514555 &    E2 &             8.3 &    1.12 &  15.17 &      0.81 &     \nodata &     \nodata &      \nodata &  \nodata &   \nodata &   \nodata &      NS &   3.1 \\
 J163910.2--514555 &  E2-1 &             8.3 &    1.12 &  15.17 &      0.81 &   18.64 &   16.28 &      \nodata &  \nodata & 0.66G & 0.34G &      CV &  10.6 \\
 J163901.5--514054 &  E4-2 &             3.7 &    2.48 &   2.93 &      0.63 &   19.75 &     \nodata &      \nodata &  \nodata & 0.19G & 0.60G &      CV &   4.5 \\
 J163906.8--515230 &  E5-1 &             3.7 &    2.66 &   3.12 &     -0.44 &   18.56 &   16.08 &      \nodata &  \nodata & 0.17G & 0.83G &      CV &   3.6 \\
 J163906.8--515230 &  E5-2 &             3.7 &    2.66 &   3.12 &     -0.44 &     \nodata &     \nodata &    13.43 &  \nodata & 0.53C & 0.47C &     AGN &   2.5 \\
 J163907.0--515018 &  E6-1 &             3.5 &    1.85 &   2.66 &      0.85 &   20.93 &     \nodata &      \nodata &  \nodata & 0.42G & 0.58G &      CV &  10.1 \\
 J163859.6--514319$^*$ &  E7-1 &             3.0 &    2.09 &   8.52 &     -1.00 &   11.21 &    9.79 &     8.93 &    $940_{-11}^{+15}$ & 0.25G & 0.54G & LM-STAR &   4.6 \\
 J163932.2--514620$^*$ &    E8 &             3.0 &    1.33 &   2.44 &      0.53 &     \nodata &     \nodata &      \nodata & \nodata &     1 &     0 &    LMXB &   2.3 \\
 J120419.8--623718$^*$ &  F7-1 &             3.2 &    2.34 &   2.20 &     -0.98 &   13.19 &   11.96 &    11.12 &     $480_{-3}^{+3}$ & 0.20G & 0.34G & LM-STAR &   2.0 \\
 J120419.8--623718$^*$ &  F7-2 &             3.2 &    2.34 &   2.20 &     -0.98 &   18.08 &     \nodata &      \nodata & $2076_{-450}^{+660}$ & 0.17G & 0.34G &    LMXB &   2.6 \\
 J120419.8--623718$^*$ &  F7-3 &             3.2 &    2.34 &   2.20 &     -0.98 &   20.34 &     \nodata &      \nodata &  \nodata & 0.14G & 0.34G &    LMXB &   2.8 \\
 J120422.3--624202$^*$ &    F8 &             3.2 &    1.28 &   1.65 &      0.48 &     \nodata &     \nodata &      \nodata &  \nodata &   \nodata &   \nodata &    LMXB &   2.8 \\
 J203443.7+363002 &    H1 &             5.7 &    1.10 &   7.71 &      0.72 &     \nodata &     \nodata &      \nodata &  \nodata &   \nodata &   \nodata &      NS &   3.4 \\
 J203443.7+363002 &  H1-1 &             5.7 &    1.10 &   7.71 &      0.72 &     \nodata &     \nodata &    15.59 &  \nodata & 0.91C & 0.09C &     AGN &  18.8 \\
 J204023.6+473315 &  I1-1 &             6.9 &    2.17 &  16.48 &      0.66 &     \nodata &     \nodata &    15.76 &  \nodata & 0.78C & 0.22C &     AGN &   2.1 \\
 J204112.7+473012 &    I5 &             3.2 &    2.29 &   1.67 &      0.19 &     \nodata &     \nodata &      \nodata &  \nodata &     1 &     0 &    LMXB &   4.4 \\
 J133000.9--610127 &    K1 &            13.9 &    0.95 &  49.54 &      0.78 &     \nodata &     \nodata &      \nodata &  \nodata &   \nodata &   \nodata &      NS &   3.1 \\
 J133000.9--610127 &  K1-1 &            13.9 &    0.95 &  49.54 &      0.78 &     \nodata &     \nodata &    13.85 &  \nodata & 0.94C & 0.06C &     AGN &   8.5 \\
 J133035.1--611121 &    K2 &             8.8 &    1.04 &  18.84 &      0.32 &     \nodata &     \nodata &      \nodata &  \nodata &   \nodata &   \nodata &      NS &   2.1 \\
 J132939.1--610744 &  K4-1 &             4.7 &    0.98 &  14.68 &     -1.00 &    9.02 &    8.13 &     7.61 &       $84_{-0.1}^{+0.2}$ & 0.72G & 0.28G & LM-STAR &   3.7 \\
 J133006.8--611555 &    K5 &             4.7 &    2.52 &   5.44 &      0.06 &     \nodata &     \nodata &      \nodata &  \nodata &   \nodata &   \nodata &    LMXB &   7.1 \\
 J133006.8--611555 &  K5-1 &             4.7 &    2.52 &   5.44 &      0.06 &   14.03 &   11.86 &    10.93 &   $2131_{-80}^{+88}$ & 0.39G & 0.61G & LM-STAR &   6.3 \\
 J133010.6--610707$^*$ &    K6 &             4.5 &    0.94 &   4.48 &     -0.65 &     \nodata &     \nodata &       &  \nodata &   \nodata &   \nodata &    LMXB &   2.3 \\
 J133010.6--610707$^*$ &  K6-1 &             4.5 &    0.94 &   4.48 &     -0.65 &   10.98 &    9.95 &     9.57 &      $370_{-4}^{+4}$ & 0.84G & 0.16G & LM-STAR &  37.1 \\
 J132928.5--611103$^*$ &  K8-1 &             4.3 &    1.40 &   3.35 &     -0.07 &   14.59 &   12.53 &    11.41 & $2143_{-183}^{+224}$ & 0.69G & 0.31G & LM-STAR &   3.8 \\
 J132947.1--611024$^*$ &  K9-1 &             4.2 &    1.01 &   3.55 &      0.33 &     \nodata &     \nodata &    15.57 &  \nodata & 0.90C & 0.10C &     AGN &   2.0 \\
 J132924.2--610417 & K12-1 &             3.4 &    2.25 &   2.84 &      0.12 &   19.57 &     \nodata &      \nodata &  \nodata & 0.17G & 0.43G &      CV &   3.0 \\
 J132924.2--610417 & K12-2 &             3.4 &    2.25 &   2.84 &      0.12 &   19.69 &     \nodata &      \nodata &  \nodata & 0.14G & 0.43G &      CV &   2.3 \\
 J132924.2--610417 & K12-4 &             3.4 &    2.25 &   2.84 &      0.12 &   19.29 &     \nodata &      \nodata &  \nodata & 0.12G & 0.43G &      CV &   2.9 \\
 J133002.5--611051$^*$ &   K13 &             3.3 &    1.17 &   1.89 &     -0.60 &     \nodata &     \nodata &      \nodata &  \nodata &   \nodata &   \nodata &    LMXB &   2.6 \\
 J133002.5--611051$^*$ & K13-1 &             3.3 &    1.17 &   1.89 &     -0.60 &   12.76 &   11.77 &      \nodata &      $796_{-8}^{+8}$ & 0.76G & 0.24G & LM-STAR &   3.9 \\
 J133018.3--610629 &   K15 &             3.2 &    1.33 &   3.29 &     -0.88 &     \nodata &     \nodata &      \nodata &  \nodata &   \nodata &   \nodata &      NS &   2.5 \\
 J133018.3--610629 & K15-1 &             3.2 &    1.33 &   3.29 &     -0.88 &   10.89 &    9.14 &     8.36 &    $886_{-24}^{+30}$ & 0.75G & 0.25G & LM-STAR &  87.9 \\
 J132935.6--611106 &   K16 &             3.1 &    1.57 &   1.61 &      0.49 &     \nodata &     \nodata &      \nodata & \nodata &     1 &     0 &      NS &   3.1 \\
 J074352.2--252403 &  L1-1 &             4.6 &    1.08 &   4.39 &     -0.34 &   13.71 &   12.07 &    11.37 &   $2001_{-56}^{+48}$ & 0.92G & 0.08G & LM-STAR &   3.4 \\
 J074428.2--252744 &    L2 &             3.8 &    2.04 &   3.27 &      0.08 &     \nodata &     \nodata &      \nodata &  \nodata &     1 &     0 &      NS &   2.3 \\
 J074346.4--252459 &    L3 &             3.5 &    1.48 &   2.48 &      0.65 &     \nodata &     \nodata &      \nodata &  \nodata &   \nodata &   \nodata &      NS &   3.7 \\
 J074346.4--252459 &  L3-1 &             3.5 &    1.48 &   2.48 &      0.65 &   21.41 &     \nodata &      \nodata &  \nodata & 0.88G & 0.12G &      CV &   3.8 \\
 J074408.8--252357 &    L4 &             3.3 &    1.11 &   1.90 &      0.50 &     \nodata &     \nodata &      \nodata &  \nodata &   \nodata &   \nodata &      NS &   3.4 \\
 J074408.8--252357 &  L4-1 &             3.3 &    1.11 &   1.90 &      0.50 &     \nodata &     \nodata &    16.08 &  \nodata & 0.88C & 0.12C &     AGN &  19.0 \\
 J135857.8--602736 &  M1-1 &             9.5 &    0.62 &  19.14 &      0.86 &   15.78 &   11.74 &      \nodata & $2311_{-181}^{+254}$ & 0.95G & 0.05G &     YSO &   3.1 \\
 J135802.9--603022 &    M3 &             4.4 &    1.29 &   8.20 &     -0.94 &     \nodata &     \nodata &      \nodata &  \nodata &   \nodata &   \nodata &      NS &   5.9 \\
\enddata
\tablecomments{ 
Variable sources    
(variability probability $>95$\%) are marked by an asterisk next to their CXO names. } 
\end{deluxetable*}

\begin{deluxetable*}{llcccccccc}
\tabletypesize{\scriptsize}
\tablecaption{Summary of the Properties for Eight $\gamma$-Ray Pulsars Coincident with 4FGL Sources 
\label{tab:4FGLDR3-NSs}}
\tablehead{
\colhead{PSR name} & \colhead{4FGL name} & \colhead{R.A.}  & \colhead{Decl.} & \colhead{P$^a$}  &  \colhead{$\dot{E}$} & \colhead{$\tau_{\rm c}$}  & \colhead{$F_{\rm b}^{\rm b}$} & \colhead{$F_{\rm b}/G_{100}$} & \colhead{References} \\
 & &  & & (ms) & ($10^{33}$ ergs s$^{-1}$) & (kyr) & ($10^{-14}$cgs) & ($10^{-4}$) &  } 
%\\
\startdata
 J1906+0722 & J1906.4+0723 & 19:06:31.20(1) &+07:22:55.8(4) & 111.5 &  1020 &  49.2 & 3.4 & 3.4 & (1), (2) \\
 J1105--6037 & J1104.9--6037 & 11:05:00.48(4) & --60:37:16.3(3) &  194.9  &  116 & 141 & 0.25 & 0.67& (1), (3) \\
J1736--3422 & J1736.1--3422 & 17:36:10.9 & --34:22:19 & 347.2 & 61 &  84 & $<0.6$ & $<3.3$ & (1) \\
J1203--6242 & J1203.9--6242 & 12:03:55.35 & --62:42:44.8 & 100.6  & 1.7$\times10^3$ &  35.8 & $<1.0$ & $<2.6$ &  (1) \\
J2034+3632 & J2035.0+3632 & 20:35:0.0 & +36:32:21 & 3.7 & 1.4 & 3.34$\times10^7$ &  $<0.4$ & $<3.3$ &  (1) \\
J0744--2525 & J0744.0--2525 & 07:44:05.5 & --25:25:17 & 92.0 & 47 & 1566 & $<0.6$ & $<3$ &  (1) \\
J1358--6025 & J1358.3--6026 & 13:58:25.88 & --60:25:55.46 & 60.5  & 535 &  319 & 1.48 & 4.8 & (1) \\
J1306--6043 & J1306.3--6043 & 13:06:20.2027 & --60:43:47.4999 & 5.67  & \nodata &  \nodata & \nodata & \nodata & (1), (4) \\
\enddata
\tablecomments{
$^a$ Spin periods of $\gamma$-ray pulsars. $^b$ When the X-ray source is not detected, we report the upper limit of the observed (absorbed) X-ray flux in the 0.5--7\,keV energy band at the 95\% confidence level. 
References: (1) \cite{2023arXiv230711132S}; (2) \cite{2015ApJ...809L...2C}; (3) \cite{2017ApJ...834..106C}; (4) \cite{2023MNRAS.524.1291P}.}
\end{deluxetable*}

\begin{deluxetable*}{llllcllccc}
\tabletypesize{\scriptsize}
\tablecaption{Summary of the Confident Classifications and X-Ray Properties (Including Spectral Fitting) of X-Ray Sources with $F_{\rm b}\ge 10^{-13}$\,erg\,s$^{-1}$\,cm$^{-2}$  and the Number of Net Counts $N_c>60$.  
\label{tab:X-ray-Class-bright}}
\tablehead{
\colhead{CXO name} & \colhead{Class} &  \colhead{PU} & \colhead{S/N}  & \colhead{$F_{\rm b}$} & \colhead{$N_c$} & \colhead{HR$_{\rm hms}$}   & \colhead{$N_{\rm H}$} & \colhead{$\Gamma$} & \colhead{C-stat/d.o.f.} \\
 & & (arcsec) & & ($10^{-13}$\,cgs) & & & ($10^{22}$ cm$^{-2}$) & &  } 
%\\
\startdata
J163830.6--514243  & 1AGN,1NS & 2.13 & 9.0 & 2.83 & 83 & 1.00 &  13$^{+7}_{-5}$ & 1.6$^{+1.0}_{-0.9}$ & 77.5/77 \\
J163910.2--514555 & 1CV,1NS & 1.12 & 8.3 & 1.52 & 70 & 0.81 &  3.1$^{+1.3}_{-1.1}$ & 2.3${\pm0.6}$ & 50.5/67 \\
J133000.9--610127 & 1AGN,1NS & 0.95 & 13.9 & 4.95 & 194 & 0.78 &  3.0$\pm0.8$ & 2.00$^{0.4}_{-0.3}$ & 121.0/146 \\
J133035.1--611121 & 1NS & 1.04 & 8.8 & 1.88 & 78 & 0.32 &  $<1.3$ & 1.4$^{+0.5}_{-0.4}$ & 52.4/60  \\
J135857.8--602736 & 1YSO & 0.62 & 9.5 & 1.91 & 92 & 0.86 & 2.9$^{+1.3}_{-1.2}$ & 1.8$\pm0.5$ & 60.7/86 \\
\enddata
\tablecomments{All X-ray sources listed in this table are not significantly variable (the variability probability is $<95\%$). }
\end{deluxetable*}

%%%%%%%%%%%%%%%%%%%%%%%%%%%%%%%%%%%%%%%%%%%%%%%%%%%%%
\bibliography{references}{}
\bibliographystyle{aasjournal}

\end{document}